\RequirePackage{fix-cm}
\RequirePackage{rotating}

\documentclass[smallextended]{svjour3}       
\smartqed  
\usepackage[lflt]{floatflt} 
\usepackage{graphicx}
\usepackage{multirow}

\usepackage[dvipsnames]{xcolor} 
\usepackage{soul} 
\sethlcolor{pink}

\usepackage{changes} 

\usepackage[title]{appendix} 

\usepackage{booktabs} 
\usepackage{soulutf8} 
\usepackage{rotating} 
\usepackage[utf8]{inputenc} 
\usepackage{multirow} 
\usepackage{colortbl} 
\usepackage{framed} 
\usepackage{dirtytalk}

\usepackage[resetlabels,labeled]{multibib}
\newcites{P}{Primary Studies}

\begin{document}

\title{A Family of Experiments on Test-Driven Development
}


\author{Adrian Santos \and
		Sira Vegas \and
        Oscar Dieste \and
        Fernando Uyaguari \and
        Ay\c{s}e Tosun \and
        Davide Fucci \and  
        Burak Turhan \and
        Giuseppe Scanniello \and
        Simone Romano \and
        Itir Karac \and        
        Marco Kuhrmann \and 
        Vladimir Mandi\'c \and
        Robert Rama\v{c} \and
        Dietmar Pfahl \and
        Christian Engblom \and
        Jarno Kyykka \and
        Kerli Rungi \and
        Carolina Palomeque \and
        Jaroslav Spisak \and
        Markku Oivo \and    
        Natalia Juristo 
}


\institute{A. Santos, I. Karac and M. Oivo \at 
           University of Oulu, Finland \\ 
           \email{\{adrian.santos.parrilla,itir.karac,markku.oivo\}@oulu.fi}
           \and
           S. Vegas, O. Dieste and N. Juristo \at
              Universidad Politécnica de Madrid, Madrid, Spain  \\
              \email{\{svegas,odieste,natalia\}@fi.upm.es}     
           \and
           F. Uyaguari and C. Palomeque \at
              ETAPA, Ecuador 
           \email{\{fuyaguar,cpalomeq\}@etapa.net.ec}           
           \and
           A. Tosun \at
              Istanbul Technical University, Turkey (University of Oulu when the experiments were run) \\ 
           \email{tosunay@itu.edu.tr}
           \and
           B. Turhan \at
              University of Oulu, Finland and Monash University, Australia \\ 
           \email{burak.turhan@oulu.fi}         
           \and
           D. Fucci \at
              Blekinge Institute of Technology, Sweden (University of Oulu when the experiments were run) \\ 
           \email{davide.fucci@bth.se}      
           \and
           G. Scanniello\at
              University of Basilicata, Italy \\ 
           \email{giuseppe.scanniello@unibas.it}
           \and
           S. Romano \at
              University of Bari, Italy (University of Basilicata when the experiments were run) \\ 
           \email{simone.romano@uniba.it}
           \and
           M. Kuhrmann \at
              University of Passau, Germany (University of Southern Denmark when the experiments were run) \\ 
           \email{kuhrmann@fim.uni-passau.de}
           \and
            V. Mandi\'c and R. Rama\v{c} \at
              University of Novi Sad, Serbia \\ 
           \email{\{vladman,ramac.robert\}@uns.ac.rs}
            \and         
            D. Pfahl \at
              University of Tartu, Estonia  \\ 
           \email{dietmar.pfahl@ut.ee}
           \and
           C. Engblom and J. Kyykka \at
              Ericsson, Finland. C. Engblom is retired now 
           \and
           K. Rungi \at
              Testlio, Estonia (PlayTech when the experiments were run)
           \and
           J. Spisak \at
              Paf, Finland \\ 
}


\maketitle

\begin{abstract}
\textit{Context}: Test-driven development (TDD) is an agile software development approach that has been widely claimed to improve software quality. However, the extent to which TDD improves quality appears to be largely dependent upon the characteristics of the study in which it is evaluated (e.g., the research method,  participant type, programming environment, etc.). The particularities of each study make the aggregation of results untenable. 
\textit{Objectives}: The goal of this paper is to: increase the accuracy and generalizability of the results achieved in isolated experiments on TDD, provide joint conclusions on the performance of TDD across different industrial and academic settings, and assess the extent to which the characteristics of the experiments affect the quality-related performance of TDD.  
\textit{Method}: We conduct a family of 12 experiments on TDD in academia and industry. We aggregate their results by means of meta-analysis. We perform exploratory analyses to identify variables impacting the quality-related performance of TDD. 
\textit{Results}: TDD novices achieve a slightly higher code quality with iterative test-last development (i.e., ITL, the reverse approach of TDD) than with TDD. The task being developed largely determines quality. The programming environment, the order in which TDD and ITL are applied, or the learning effects from one development approach to another do not appear to affect quality. The quality-related performance of professionals using TDD drops more than for students. We hypothesize that this may be due to their being more resistant to change and potentially less motivated than students.
\textit{Conclusion}: Previous studies seem to provide conflicting results on TDD performance (i.e., positive vs. negative, respectively). We hypothesize that these conflicting results may be due to  different study durations, experiment participants being unfamiliar with the TDD process, or case studies comparing the performance achieved by TDD vs. the control approach (e.g., the waterfall model), each applied to develop a different system. Further experiments with TDD experts are needed to validate these hypotheses. 
\
\end{abstract}

\section{Introduction}
\label{sec:introduction}

Test-driven development TDD is an agile software development approach stipulating the construction of software systems by means of micro iterative testing-coding cycles---as opposed to test-last approaches, where coding and testing are rarely interleaved, and coding is usually performed before testing. According to TDD proponents \cite{astels2003test,beck2003test}, these micro-iterative testing-coding cycles are the main reason why TDD outperforms test-last development approaches in terms of software quality.  Although the TDD literature \cite{bissi2016effects,causevic2011factors,kollanus2010test,makinen2014effects,munir2014considering,rafique2013effects,shull2010we} examines several quality attributes, we focus here on \textit{external quality}. External quality is one of the most researched attributes, which, according to its proponents \cite{astels2003test,beck2003test}, TDD benefits the most. It is usually considered in the TDD literature as the number of successful test cases in a test battery (i.e., test oracle \cite{bertolino2007software}) specifically built for testing the application under development \cite{causevic2011factors,kollanus2010test,rafique2013effects,shull2010we}\footnote{For simplicity's sake, we refer to quality and external quality interchangeably throughout the rest of the article. We acknowledge the limitations of this under the threats to validity.}. In this article, we define quality as the percentage of test battery tests developed by the experimenters that pass. 

The rationale of TDD proponents is seemingly intuitive: if tests are built up-front in small iterations---rather than after the code has been developed, that is, \textit{\say{\ldots Don't write a line of new code unless you first have a failing automated test\ldots}} \cite{beck2003test}---the developers put a limit on the code \cite{astels2003test}, think up-front about its design \cite{beck2003test}, and reduce their stress in response to potential errors \cite{beck2003test}. These continuous and incremental iterations lead to a \textit{virtuous cycle} \cite{beck2003test} that eventually translates into \textit{\say{\ldots simpler designs\ldots, systems that reveal intent\ldots, [and] extremely low-defect systems that start out robust, are robust at the end, and stay robust all the time\ldots}} \cite{astels2003test}. 

Although this may sound too good to be true, there is no scarcity of empirical studies supporting these claims \cite{bissi2016effects,causevic2011factors,kollanus2010test,makinen2014effects,munir2014considering,rafique2013effects,shull2010we}. However, the extent to which TDD improves quality is still inconclusive, especially in view of the large heterogeneity of results reported in the literature, and the many dimensions on which the studies differ. As many contextual factors vary across the published studies, we wonder what impact they may be having on the reported results.

Families of experiments---a term coined by Basili et al. in 1999 \cite{basili1999building} for referring to groups of interrelated experiments pursuing the same goal---are on the rise in software engineering (SE) \cite{adrisms}. However, there is no consensus on how different interrelated experiments can be. Gomez et al. \cite{gomez2014understanding} state that they can lie: from identical replications (i.e., experiments following exactly the same procedures and operationalizations as baseline experiments), to conceptual replications (i.e., experiments that only share research questions with baseline experiments). Some other authors argue that they should only be conceptual replications (e.g., Kitchenham \cite{kitchenham2008role}). In our experience after performing a Systematic Mapping Study (SMS) to identify the families of experiments already published in SE \cite{adrisms}, we have seen that experiments typically share experimental design, response variable operationalizations, and research questions and objectives. Nevertheless, some slight changes are usually made across the replications (e.g., in terms of experimental session length or participant type evaluated).

Albeit not without shortcomings (e.g., families tend to be composed of fewer studies than are usually gathered in systematic literature reviews (SLRs), families usually study fewer response variables than SLRs, etc. \cite{adrisms}) and have some strengths that make them good for assessing the performance of SE interventions (such as TDD) \cite{adrisms}:
\begin{itemize}
    \item{As families do not rely on already published studies, families avoid the potential effects of publication bias (i.e., the tendency to publish more favorable than unfavorable results \cite{cooper2009relative}) on joint outcomes.}
    \item{As there is access to raw data within families (since the researchers run the experiments themselves), it is possible to apply consistent procedures for analyzing each experiment individually. This ensures that differences across experiment results are not due to the use of different procedures to analyze each experiment, but, instead, due to real differences in the collected data \cite{cooper2009relative}.}    
    \item{As the response variables can be measured identically in all the experiments, families avoid the potentially detrimental effects of combining the results of studies with different response variable operationalizations into joint conclusions.}
    \item{As the researchers conducting families are able to design the experiments as desired, they can systematically vary the conditions of the experiments (e.g., their programming languages, experimental designs, etc.) with the aim of assessing the influence of such changes on results \cite{juristo2009using}, or fixing identified flaws in the design}.
\end{itemize}

To the best of our knowledge, the Experimental Software Engineering Industry Laboratory (ESEIL Project) is the first research project undertaking a family of experiments on TDD across several software industrial partners and universities. With our family, we aim to answer three \textbf{research questions}:
\begin{itemize}
    \item{\textbf{RQ1:} Do TDD and iterative test-last development (ITL) perform similarly in terms of quality?}
    \item{\textbf{RQ2:} To what extent is quality affected by the task under development?}    
    \item{\textbf{RQ3:} To what extent is quality affected by the experiment characteristics: participant type (i.e., professionals vs. students), programming environment\footnote{Note that in our experiments the programming language is confounded with other variables: IDE, testing tools, and other programming environment related variables (the use of Java implies the use of Java-related technologies, while the use of C++/C\# implies the use of C++/C\#-related technologies). We have grouped all confounded variables under the \textit{programming environment} name.} (i.e., Java and related IDEs and testing tools vs. C++/C\# and related IDEs and testing tools), learning effects from one session to another, or order of application of ITL and TDD?}
    \item{\textbf{RQ4:} To what extent is quality affected by the participant characteristics: programming, programming language, unit testing, or testing tool experience?}

\end{itemize}

To answer these research questions, we conducted 12 controlled experiments in both industry and academia. In all the experiments, we measured quality identically and had the participants code identical tasks. We followed a mixed approach for designing the experiments: while we adapted the programming environments and experimental designs to the requirements of our industrial partners, we explored the potential limitations of these experiments by comparing their results with the outcomes of the academic experiments that we ran. We followed a two-step data analysis approach: we first conducted a meta-analysis to output joint results \cite{borenstein2011introduction} (answering RQ1), and then we performed a series of exploratory analyses (i.e., sub-group meta-analyses \cite{borenstein2011introduction} and linear mixed models \cite{brown2014applied}) to assess the influence of third variables on results (answering RQ2 and RQ3). Along the way, we made several \textbf{findings}:
\begin{itemize}
    \item{TDD novices (both professionals and students) perform slightly better with ITL than with TDD in terms of quality---although the difference in performance is not statistically significant. Our results suggest that not much improvement would be achieved with the application of TDD over ITL by TDD novices.}
    \item{TDD performance appears to drop more for professionals than for students (compared to performance using ITL). This suggests either that students learn completely new development approaches (such as TDD) faster than professionals or that the drop in quality for professionals applying TDD is greater than for students because professionals achieve higher quality than students when using ITL.}
    \item{Programming environment (i.e., Java environment, or C++/C\# environment), the order in which TDD is applied (i.e., in the first or second session), the learning effects from one session to the next, or the task being developed---as long as the same task is developed with both TDD and ITL---do not appear to have an impact on the difference in performance between TDD and ITL. }
\end{itemize}

The main \textbf{contributions} of this paper are \textit{an assessment of the quality-related performance of TDD in a family of experiments} and a \textit{series of exploratory analyses} to evaluate the extent to which the experiment and participant characteristics affect TDD performance in industrial and academic settings. We think this article rounds out previous research in several ways:

\begin{enumerate}

	\item This is the first family of experiments on TDD. Besides, this is one of the largest families in SE research with respect to number of experiments (12),  universities (8), companies (4), and participants (411) \cite{santos2018profes}. This research should increase the reliability of the findings, and the generalization of results to different contexts and populations \cite{basili1999building}.

	\item This family of experiments is one of the few studies on TDD---apart from \cite{rafique2013effects}---that attempt to provide joint conclusions by means of meta-analysis. This should increase the reliability of the joint conclusions with respect to findings reached using a narrative approach of counting positive vs. negative results \cite{borenstein2011introduction,gurevitch2018}. The narrative  approach is typically adopted in secondary and tertiary studies on TDD---perhaps because it is not feasible to apply meta-analysis on the studies published in the literature.
	
	\item This family of experiments includes the results of four industrial experiments on TDD. The need for more industrial experiments on TDD was already highlighted by Munir et al. \cite{munir2014considering} and Rafique and Mi\v{s}i\'c \cite{rafique2013effects}. 

	\item This family of experiments compiles the primary studies identified in previous secondary studies, remaps and organizes them with regard to their characteristics and results, and provides further lines of research in view of both the results of our family of experiments, and the resulting map. This adds to the work of others, calling upon  the community to conduct more research on TDD \cite{offutt2018don}.

	\item This study is the first on TDD that aims to study the influence of participant characteristics on results across multiple experimental settings and populations. We can do this as we have access to the raw data of all the experiments and also because we used similar instruments to measure the participant characteristics and external quality across the experiments. This may be unfeasible in SLRs: either because the raw data are not available, or because different measurement instruments are used across the published experiments to measure participant experience.

	\item This family of experiments is one of the few SE families that provide the raw data and R code to make results reproducible \cite{santos2018profes}. Also, it is one of the few families planning to make experimental data available (i.e., the code of participants). This should facilitate re-analysis, the application of different analysis methods to provide joint conclusions, the investigation of other research questions, and the integration of results in future meta-analyses. 
	
	\item The results of this family of experiments contradict findings commonly reached in secondary and tertiary studies on the effectiveness of TDD on external quality \cite{karac2018we}. While secondary and tertiary studies typically indicate that TDD achieves higher quality than control approaches \cite{karac2018we}, we do not reach the same conclusion. Also, our results disagree with the findings of the meta-analyses conducted by \cite{rafique2013effects} claiming that industrial experiments led to more significant improvements in quality with TDD. In particular, we reached the opposite conclusion in the industrial experiments that we ran (i.e., the performance of professionals with TDD dropped to a larger extent than for students). The difference in results could be due to the control used. In our family of experiments, we use ITL (instead of the waterfall model) as a control. The only difference between ITL and TDD is when tests are generated (either before or after coding), but both use short development cycles. The waterfall model implies that tests are generated after coding and there are no development cycles. The benefits of TDD could be due to the use of these short development cycles (which is also a characteristic of ITL \cite{karac2018we}, but not of the waterfall model). In view of this, we call upon the community not only to look at the results of the studies but to also assess the extent to which participant or study characteristics may be influencing results.
	
	\item This family of experiments partially supports the results of \cite{fucci2017dissection}, one of the latest publications on TDD in testing software engineering, namely, that the order of testing (either first, as in TDD, or last, as in ITL) does not appear to influence external quality. Indeed, although ITL achieved slightly higher quality than TDD in our family of experiments, the difference was not relevant. What really appears to be improving quality is not when testing is performed---first or last---but rather the average duration of development cycles, the duration uniformity and the refactoring effort.
	
	\item This family of experiments partially supports Beck's suggestions \cite{beck2003test} that \textit{virtuous cycles}---leading to higher performance---materialize with TDD. However, such \textit{virtuous cycles} may only materialize with students that have quite a lot more programming experience than their peers. This is a ground for further research to double-check our observations.

\end{enumerate}

\textbf{Paper organization}. Section \ref{related_work} discusses the work related to our study. In Section \ref{family_design}, we describe the characteristics of our family of experiments on TDD. Then, Section \ref{rq1}, Section \ref{rq2}, Section \ref{rq3} and Section \ref{rq4} answer  RQ1, RQ2, RQ3, and RQ4, respectively, and discuss the implications of our findings. In Section \ref{threats_to_validity}, we outline the threats to validity of our study. Finally, we outline our conclusions in Section \ref{conclusions}.

\section{Related Work}
\label{related_work}
  
\begin{sidewaystable*} \centering 
  \caption{Primary studies assessing the quality-related performance of TDD.} 
  \label{table_related_work} 
\begin{tabular}{cclllllll} \hline \hline 
\textbf{ID} & \textbf{Result (\%)} & \textbf{Method} & \textbf{Control} & \textbf{Participant} & \textbf{Task} & \textbf{Unit} & \textbf{Length} & \textbf{Environment} \\ 
\hline 
\citeP{vu2009evaluating} & -39 & Case study & Waterfall & BSc & Industrial & Team & Months & Google Web Kit, Adobe Flex, SVN \\ 
\citeP{zielinski2005preliminary} & -33 & Experiment & Waterfall  & MSc & Toy & Solo & Weeks & Java,Eclipse, CVS \\ 
\citeP{mueller2002experiment} & -30 & Experiment & Waterfall  & MSc & Toy & Solo & Days & Java, JUnit, Unix \\ 
\citeP{wilkerson2012comparing} & -5 & Case study & Waterfall  & BSc & Toy & Solo & Weeks & Java, JUnit, Eclipse \\ 
\citeP{huang2009empirical} & -4 & Case study & Waterfall  & BSc & Industrial & Team & Months & Java, PHP, Eclipse, JUnit, CVS \\ 
\citeP{erdogmus2005effectiveness} & -3 & Case study & ITL & BSc & Toy & Solo & Days & Java, JUnit, Eclipse, CVS \\ 
\citeP{pancur2003towards} & -3 & Case study & ITL & BSc & Toy & Pairs & Weeks & Java, JUnit, Eclipse \\ 
\citeP{geras2004prototype} & 0 & Case study & Waterfall  & Prof & Toy & Team & Days & JUnit, VBExpect; SQL \\ 
\citeP{lui2004test} & 2 & Case study & Waterfall  & Prof & Industrial & Team & ? & ? \\ 
\citeP{panvcur2011impact} & 2.5 & Experiment & ITL & BSc & Toy & Solo & Days & Java, JUnit, Eclipse \\ 
\citeP{desai2009implications} & 3 & Case study & Waterfall  & BSc & Toy & Solo & Months & Java, JUnit \\ 
\citeP{xu2009evaluation} & 11 & Experiment & Waterfall  & BSc & Toy & Solo & Days & Java; JUnit, Eclipse \\ 
\citeP{gupta2007experimental} & 12 & Experiment & Waterfall  & BSc, MSc & Toy & Solo & Weeks & Java, JUnit, DrJava editor \\ 
\citeP{domino2007controlled} & 12 & Experiment & Waterfall  & BSc, MSc, Prof & Toy & Solo & Days & Pseudocode \\ 
\citeP{george2004structured} & 18 & Experiment & Waterfall  & Prof & Toy & Pairs & Days & Java, JUnit \\ 
\citeP{edwards2003using} & 22 & Case study & Waterfall  & BSc & Toy & Solo & Weeks & Java, JUnit \\ 
\citeP{rahman2007applying} & 24 & Case study & Waterfall  & BSc & Toy & Solo & ? & ? \\ 
\citeP{yenduri2006impact} & 36 & Case study & Waterfall  & BSc & Toy & Solo & Months & Java \\ 
\citeP{slyngstad2008impact} & 36 & Case study & Waterfall  & Prof & Industrial & Team & Years & Java \\ 
\citeP{sanchez2007sustained} & 40 & Case study & Waterfall  & Prof & Industrial & Team & Years & Java, ANT \\ 
\citeP{maximilien2003assessing} & 50 & Case study & Waterfall  & Prof & Industrial & Team & Years & Java, JUnit, Ant \\ 
\citeP{damm2007quality} & 55 & Case study & Waterfall  & Prof & Industrial & Team & Years & C++, XML \\ 
\citeP{george2002analysis} & 62 & Experiment & Waterfall  & BSc & Toy & Pairs & Days & Java, JUnit \\ 
\citeP{dogvsa2011effectiveness} & 80 & Case study & Waterfall  & Prof & Industrial & Team & Months & JUnit,  Java, Netbeans, CVS \\ 
\citeP{nagappan2008realizing} & 90 & Case study & Waterfall  & Prof & Industrial & Team & Years & Java, JUnit, C, C++, C\# \\ 
\citeP{madeyski2007impact} & 96 & Case study & ITL  & Prof & Toy & Solo & Days & Java, AspectJ, JSP, XML, Eclipse \\ 
\citeP{bhat2006evaluating} & 100 & Case study & Waterfall   & Prof & Industrial & Team & Years & C/C++, CUnitTest \\ 
\citeP{ynchausti2001integrating} & 267 & Case study & Waterfall  & Prof & Toy & Pairs,  Solo & Days & ? \\ 
\citeP{bannerman2011multiple} & - & Case study & Waterfall  & Prof & Industrial & Team & Years & Java, Ant, CVS, SVN \\ 
\citeP{kobayashi2006analysis} & - & Case study & - & Prof & Industrial & Team & Months & C\#, Java, Struts \\ 
\citeP{siniaalto2008does} & - & Case study & ITL & MSc, Prof & Industrial & Team & Months & Java \\ 
\citeP{marchenko2009long} & - & Case study & Waterfall  & Prof & Industrial & Pairs & Years & Java \\ 
\citeP{aniche2010most} & - & Survey & Waterfall  & Prof & - & - & Weeks & - \\ 
\citeP{lejeune2006teaching} & - & Case study & Waterfall  & BSc & Toy & Team & Weeks & Java, JUnit, CVS \\ 
\citeP{mcdaid2008test} & - & Case study & Waterfall  & Prof & Industrial & Solo & ? & Java, JUnit? \\ 
\citeP{paula2006quality} & - & Case study & - & MSc & Toy & Solo & Days & Java, JUnit, Swing \\ 
\hline
\end{tabular} 
\end{sidewaystable*}

Here we provide an overview of the research that has been conducted so far on the quality-related performance of TDD based on the results of the secondary studies addressing this question \cite{bissi2016effects,causevic2011factors,kollanus2010test,makinen2014effects,munir2014considering,rafique2013effects,shull2010we}. In particular, we first went through the secondary studies and gathered all the primary studies that---according to the authors of the secondary studies---studied external quality. Then, we removed duplicates by title and by double publication (i.e., any primary studies that had already been published in other venues), selecting the most recent publications. Finally, we extracted the following information from each primary study for the purposes of classification:

\begin{itemize}
    \item{\textit{The results achieved}: the response ratio effect size \cite{borenstein2011introduction}. First we calculated the response ratio by dividing the mean of the quality scores achieved with TDD by the mean of the quality scores achieved with the control approach (e.g., ITL or waterfall model). For example, if 30 bugs have been reported in a system developed with ITL and 10 in a system developed with TDD, the response ratio is equal to 3. We multiply the response ratio by 100\% to convey the percentage improvement of TDD over ITL. In the above example, TDD outperformed ITL by 300\%.}
    \item{\textit{The research method} according to Wohlin et. al.'s definitions \cite{wohlin2012experimentation}:
    
    \begin{itemize}
        \item{Surveys: empirical studies that collect information from or about people to describe, compare or explain their knowledge, attitudes and behavior For us, the defining characteristic of surveys is that their results are based upon the opinions of the respondents.}
        
        \item{Case studies: empirical studies that draw on multiple sources of evidence to investigate one instance of a phenomenon within its real-life context. For us, the defining characteristic of case studies is that their results are based upon those observed in uncontrolled but real-life environments.}
        
        \item{Experiments: empirical studies that investigate a certain phenomenon in a controlled---albeit scaled down---environment. For us, the defining characteristic of experiments is that all the contextual factors of the experiment (e.g., the programming environment, the tasks under development, etc.) have an identical impact on all the interventions being compared.}
    \end{itemize}}
    
    \item{\textit{The control:} the whole system coded before testing (waterfall) or the reverse approach of TDD (ITL).
    
    \item{\textit{Participant type:} we used bachelor students (BSc), master students (MSc), or professionals (Prof).}
    
    \item{\textit{Task type:} the systems developed by the participants are either toy tasks (i.e., small tasks with a reduced scope), or industrial tasks (i.e., large tasks with real-life requirements).}
    
    \item{\textit{The unit of analysis:} quality is evaluated for solo programmers (i.e., solo), for pair programmers \cite{williams2002pair} (i.e., pairs) or for groups (i.e., teams).}
    
    \item{\textit{Length of the study}: days, weeks, months, or years.}
    
    \item{\textit{Programming environment:} the technologies that may potentially interact with the performance of TDD (e.g., the programming language, the testing tool used, etc.).}}
    
\end{itemize}

Table \ref{table_related_work} shows the list of the primary studies\footnote{Due to space restrictions, we moved the references of the primary studies to the supplementary material (see Appendix A).} that we selected (ordered by response ratio, with the worst results for TDD first). Notice that:

\begin{itemize}

    \item \textit{The results of the different studies are hugely variable}. For example, TDD outperforms ITL in terms of quality measured by successful tests in a four-hour experiment with undergraduate students coding toy tasks with Java, JUnit and Eclipse \citeP{panvcur2011impact} by just 2.5\%. On the other hand, TDD outperforms the waterfall model in terms of quality measured by reported bugs in a case study at Microsoft, where a small team of developers used C++ and C\# to implement a new functionality in an industrial system over several years \citeP{nagappan2008realizing}, by up to 90\%. The variation ranges from -39\% to +267\%.  
    
    \item{\textit{In most studies, TDD achieves higher quality than control approaches} (mostly the waterfall model) both quantitatively (20 out of 36) and textually\footnote{It was not feasible to compute any response ratio synthesizing the quality achieved with TDD with respect to a control approach.} (the eight studies marked with a '-'). Detrimental effects were reported by only seven studies, of which four showed negligible effects (from -5\% to -3\%).}
   
    \item{\textit{Most primary studies are case studies} (27 out of 36). TDD was evaluated in a controlled environment in only eight experiments. }
    
    \item{\textit{Case studies report the most optimistic results for TDD}. When case studies show detrimental effects, they tend to be negligible (except for \citeP{vu2009evaluating} with undergraduate participants).}
    
    \item{\textit{Professionals tend to participate only in case studies}, which tend to evaluate TDD performance for teams rather than for solo programmers (i.e., the resulting quality of the whole system after it has been implemented by a group of developers). In the only two controlled experiments in which professionals participated and were evaluated as either solo or pair programmers (i.e., \citeP{domino2007controlled,george2004structured}), the results were not as favorable to TDD.}
    
    \item{\textit{The most beneficial results for TDD tend to be achieved for teams}. On the contrary, the performance of TDD for solo programmers shows conflicting evidence (ranging from -33\% to +96\%). In general, the studies evaluating solo or pair programmers show less optimistic results than those evaluating teams (except for the outlier in \citeP{ynchausti2001integrating}\footnote{The outlier observed in \citeP{ynchausti2001integrating} may have been due to the small number of participants, and the larger variability of results expected in small sample sizes \cite{cumming2013understanding}.}).} 
    
    \item{\textit{TDD has been evaluated with both toy tasks and industrial tasks (i.e., 16 and 20, respectively)}: Professionals are usually involved in studies evaluating the performance of TDD in industrial tasks, whereas students participate in studies with toy tasks. The studies with toy tasks tend to provide less optimistic results for TDD.}
    
    \item{\textit{The studies with longer periods of evaluation tend to report the most favorable results for TDD}. This suggests that it may be necessary to apply TDD for months or years, rather than days or weeks, to get larger benefits from this approach. }
    
    \item{\textit{TDD has been mostly evaluated with Java technologies}. However, some isolated studies have also evaluated TDD with C technologies (e.g., C/C++/C\# and related IDEs and testing tools). In such studies, the results are always favorable to TDD. }
    
\end{itemize}

Summarizing, most studies indicate that TDD is superior to control approaches (i.e., mostly the waterfall model) with respect to quality. However, the extent to which TDD achieves higher quality than control approaches appears to be largely dependent upon the characteristics of the study in which TDD is evaluated. The heterogeneity of research methods, programming environments, lengths of evaluation, units of analysis, task types, and participant types makes the aggregation of results untenable. Besides, some other issues rule out the quantitative aggregation of results, and the search for potential variables explaining the disparity of results, including:

\begin{itemize}
    \item{\textit{The different practices applied in tandem with TDD}. For example, while TDD and pair programming were applied together in some studies (e.g., \citeP{george2002analysis,kobayashi2006analysis}), other studies applied TDD either together with other agile practices (e.g., \citeP{huang2009empirical}), or merely as a part of custom-made development processes (e.g., \citeP{paula2006quality}). Thus, the aggregation of the results of these studies---each applying different practices in tandem with TDD---to provide joint conclusions would be tantamount to mixing apples and oranges \cite{borenstein2011introduction}. This may affect the reliability of joint results.}
    \item{\textit{The different ways in which quality was measured}. While quality is commonly measured as the percentage of test battery test cases that pass (e.g., \citeP{erdogmus2005effectiveness,george2004structured,george2002analysis,zielinski2005preliminary}), external quality is measured as the number of defects that were found in bug tracking systems (e.g., \citeP{bhat2006evaluating,nagappan2008realizing}) in some studies or as a customized scale as a result of applying scoring templates on the solutions of the participants (e.g., \citeP{domino2007controlled}) in others. This heterogeneity of response variable operationalizations may be an obstacle to the aggregation of results.}
\end{itemize}

In this study, we followed a different approach to others adopted in the literature with the hope of moving beyond the limitations of aggregating published results. In particular, we conducted a family of experiments (versus running one single empirical study) on TDD (versus applying TDD in tandem with other approaches) and used identical scales to measure quality across all of our experiments (versus using different metrics to measure quality). We also trained the participants across all our experiments in identical TDD and ITL procedures and had them code identical tasks. Thus we were able to use formal analysis methods to aggregate the results of the experiments (e.g., meta-analysis \cite{borenstein2011introduction} and linear mixed models \cite{brown2014applied}). We also made some changes across the experiments. As a result, we were able to assess the extent to which the characteristics of the experiments may have affected the results. Finally, assuming that the results in our family are representative of what may be happening in other studies on TDD, we were able to hypothesize on the  variables potentially behind the disparity in the results observed so far in the literature. In the following, we outline the design of our family of experiments.

\section{Family Design}
\label{family_design}

We report the variables, objectives, and the tasks developed by the participants in our family of experiments in Section \ref{variables_research_question}. We provide an overview of the characteristics of the experiments in Section \ref{design}. We focus on the ethics of the experiments in Section \ref{ethics}. We describe participants experience in Section \ref{participants_description}. Finally, we discuss the data analysis in Section \ref{analysis_approach}

\subsection{Variables, Tasks and Objectives}
\label{variables_research_question}

The main \textbf{factor} (independent variable) in our family of experiments is the \textit{development approach}, with \textit{TDD} and \textit{ITL} (following Tosun et al. \cite{tosun2017industry}) as treatments\footnote{Throughout the rest of the paper we refer to the \textit{treatment}---terminology commonly used in experimental design and data analysis \cite{brown2014applied,higgins2008cochrane,juristo2013basics,wohlin2012experimentation}---and the \textit{development approach} (i.e., either ITL or TDD) interchangeably.}. ITL is closely related to TDD in terms of constitutive elements (breaking the specification into smaller subtasks, testing, coding, and refactoring). ITL differs from TDD regarding the order of these steps (test-last versus test-first, respectively). ITL has been chosen as the control instead of the waterfall model because in the waterfall model testing is done at the end. If the result were that TDD performs better than the waterfall model, it could be because the participants have not tested, and have spent most of the time coding. This does not happen in the case of ITL, where it is guaranteed that testing is done at the end of each development cycle.

TDD and ITL were applied by the participants in our experiments to develop at least one out of five \textit{toy tasks}: MarsRover (MR), Bowling Scorekeeper (BSK), MusicPhone (MP), Sudoku (SDK) and Spread-Sheet (SS). As MP and SDK were only developed in one experiment (i.e., University 1), and SS in two experiments (i.e., Company 3 and 4), they are not reviewed here for reasons of space. However, interested readers are referred to the supplementary material where we provide all specifications\footnote{https://github.com/GRISE-UPM/FiDiPro\_ESEIL\_TDD.}.

BSK is a modified version of Robert Martin's Bowling Scorekeeper \cite{martin2001advanced}. The goal of the task is to calculate the score of a single bowling game. The task is algorithm-oriented. BSK does not require prior knowledge of bowling scoring rules; this knowledge is embedded in the specification. MR is a programming exercise that requires the development of a public interface for controlling the movement of a fictitious vehicle on a grid with obstacles. MR is a popular exercise in the agile community to teach and practise unit testing. There is a difference in the level of granularity of the task specifications. While BSK's specifications are \textit{fine-grained} (i.e., detailed specifications where the functionality of the task is divided into small chunks that can be atomically implemented one after another), MR's specifications are \textit{coarse-grained} (i.e., less detailed, and without dividing larger functionalities into smaller parts). We used the same specifications for BSK and MR in our experiments as were used in previous TDD experiments in the literature \citeP{erdogmus2005effectiveness},\cite{dieste2017empirical,fucci2017dissection,tosun2017industry}. The size of the MR and BSK programs that we---the experimenters---developed is 295 LOC and LOC respectively. We discuss the trade-offs of using toy tasks in the threats to validity section. 

The main \textbf{response variable} (dependent variable) within the family is \textit{external quality}. As usual in the TDD literature, we measure external quality as the percentage of successful tests in the test battery---or \textit{test oracle} \cite{bertolino2007software}---that we (i.e., the experimenters) built to test participant solutions. The participants had no access to these test oracles. We measured external quality as:
$$ QLTY =\frac{\#Tests(Pass)}{\#Tests(All)}*100 \%$$

We consider this to be an appropriate measure of external quality, since when measured from a functional viewpoint, it would be equivalent for both ITL and TDD.

We built a different test oracle to test each task. We used the same test oracle to measure the external quality of the code produced by all participants. When tests failed due to trivial errors in the code, we corrected it. Table~\ref{gold_standard} shows the \textit{total number of test cases} that make up the MR and BSK test oracles, testing \textit{statement coverage} and \textit{branch coverage}\footnote{Both measured with eclEmma: https://www.eclemma.org/ } \cite{myers2011art}, and \textit{mutation score}\footnote{Measured with muJava: https://cs.gmu.edu/~offutt/mujava/} \cite{jia2011analysis} for an implementation that we---the experimenters---coded for each task based on their specifications. 

\begin{table}[h!] \centering 
  \caption{Test oracles for MR and BSK.} 
  \label{gold_standard} 
\begin{tabular}{lcccc} \hline \hline 
\textbf{Task} & \textbf{Number} & \textbf{Statement} & \textbf{Branch} & \textbf{Mutation} \\ 
& \textbf{of tests} & \textbf{coverage} &  \textbf{coverage} & \textbf{score} \\ 

\hline 
MR & 52 & 100\% & 84.9\% & 82\%\\
BSK & 48 & 100\%& 96.4\% & 94\%\\ \hline
\end{tabular} 
\end{table}

Finally, we define the \textbf{objective} of our family of experiments following the Goal-Question-Metric (GQM) paradigm proposed by Basili et al. \cite{basili1992software} as follows: \\

\noindent\fbox{
  \parbox{11.5cm}{
     \textbf{GQM}. Analyze \textit{TDD and ITL} for the purpose of \textit{comparison} with respect to \textit{external quality} from the point of view of \textit{the researcher} in the context of \textit{TDD novices coding toy tasks in controlled experiments}.
  }
}

\subsection{Design of the Experiments}
\label{design}

\begin{table*}[t!]
\footnotesize
\begin{center}
\caption{Family of experiments on TDD: experimental designs (ordered by experimental design).}
\label{tab:experimental_design}
\begin{tabular}{ c| l | l | l | l| l| l} \hline \hline
\textbf{ID} & \textbf{Experiment} & \textbf{N} & \textbf{Design} & \textbf{Participant} & \textbf{Task} & \textbf{Environment} \\
& & & & \textbf{Type} &  & \\ \hline
1&University 1 & 16 & Within & Students & MR, BSK, & Java, JUnit, Eclipse \\ 
& & & & & SDK, MP & \\ \hline

2&Company 1 & 18 & Within & Professionals &  MR, BSK & Java, JUnit, Eclipse \\
3&Company 2 & 20 & Within  & Professionals& MR, BSK & C++, Boost, Eclipse, Vim  \\ 
4&University 2& 48 & Within & Students& MR, BSK & Java, JUnit, Eclipse \\
5&University 3& 53 & Within & Students&  MR, BSK &  Java, JUnit, Eclipse\\ \hline

6&Company 3& 8 & Within &Professionals & MR, BSK, & Java, JUnit, Eclipse  \\ 
& & & & & SS& \\ 

7&Company 4 & 13 & Within & Professionals & MR, BSK, &  C\#, GxUnit, GeneXus\\ 
& & & & & SS& \\ 

\hline
8&University 4 & 20 & Crossover &Students &  MR, BSK & Java, JUnit, Eclipse\\ \hline

9&University 5 & 41 &Between  &Students &   BSK & Java, JUnit, Eclipse \\
10&University 6 & 69 &Between & Students & BSK & C\#, NUnit, Visual Studio \\ 
11&University 7& 64& Between &Students & BSK & C\#, NUnit, Visual Studio  \\
12&University 8 & 41 & Between & Students & MR, BSK & Java, JUnit, Eclipse\\
\hline

\end{tabular}
\end{center}
\end{table*}

We embedded all our industrial experiments within TDD training courses so as to increase their appeal to practitioners \cite{vegas2015difficulties}, even though this approach has its own shortcomings. As experiments are embedded within training courses, the professionals attending the experiments are TDD novices. None of the participants reported having any TDD or ITL experience in our questionnaire\footnote{Note that the fact that participants do not have any ITL and TDD experience does not mean that they have no software testing experience. ITL and TDD have to do with knowledge of slicing and not with knowledge of testing. Therefore, participants with testing experience might conceivably have no experience with either ITL and/or TDD.}. The decision to embed experiments within training courses forced us to make certain design decisions. 

First, the professionals attending the training courses needed to apply both ITL and TDD---as otherwise, they might complain that they did not exercise all the development approaches taught during the course. Thus, we had to rely on within-subjects designs for all our industrial experiments. Second, as all the participants needed to exercise both development approaches, we had to make the participants code different tasks with each development approach---as otherwise, successive attempts at coding the same task could boost their performance, regardless of the development approach applied. As a result, we had to introduce more than one task in our industrial experiments.

Having these requirements in mind, we decided to design all our industrial experiments as AB within-subjects experiments (see Table \ref{clarification_ab_within}). We decided to run AB within-subjects experiments---instead of crossover experiments (see Table \ref{clarification_cross_over})---as they fitted the structure of a training course better. In particular, thanks to this design, we were able to train participants in ITL on the first day and in TDD on the last day, running the experimental session to assess the performance of each development approach immediately after each training session. Note that the companies mainly see a course. Therefore, we had to prioritize a design that was as least disturbing as possible for training purposes.

\begin{table}[h!] \centering 
  \caption{AB within-subjects design.} 
  \label{clarification_ab_within} 
\begin{tabular}{ccc} \hline \hline 
\textbf{Group} &\textbf{Day 1 (ITL)} & \textbf{Day 2 (TDD)} \\ \hline \\[-1.8ex] 
G1&Task 1 & Task 2 \\
G2&Task 2 & Task 1 \\
\hline \\
\end{tabular} 
\end{table}

\begin{table}[h!] \centering 
  \caption{Crossover design.} 
  \label{clarification_cross_over} 
\begin{tabular}{ccc} \hline \hline 
\textbf{Group} &\textbf{Day 1 (Task 1)} & \textbf{Day 2 (Task 2)} \\ \hline \\[-1.8ex] 
G1 &ITL & TDD \\
G2 &TDD & ITL \\
\hline \\
\end{tabular} 
\end{table}

Even though an AB within-subjects design suited the structure of our training courses, this design has a major shortcoming: the order in which the development approaches are applied may distort results \cite{wohlin2012experimentation}. This is because the participants may learn something during the first session (in this case, the ITL session) that could boost their performance in the second session (in this case, the TDD session). For example, they may learn how to code in short iterations in the first session, which could boost their performance with TDD. In order to assess the extent to which the order of application distorted the results, we compared the results of our AB within-subjects experiments with the outcome of a crossover experiment that we ran (see below).

We had more degrees of freedom for the design of the experiments in academia. In particular, and as these experiments were part of longer programming courses, we could run both within-subjects and between-subjects experiments. We ran between-subjects experiments (i.e., experiments in which each participant applies just one development approach, either TDD or ITL) for the experiments where we expected more participants. By running between-subjects experiments, we were able to assess the extent to which applying one development approach, instead of two, affected the results. We designed the rest of our experiments in academia as either AB within-subjects experiments or as a crossover experiment.

Finally, we adapted the programming environments to the requirements of our host intuitions. These adaptations are a double-edged sword: on the one side, they may lead to heterogeneity of results (as if the programming environments are different, the results of the experiments may also be different); on the other side, they also represent an opportunity to increase the external validity of the results (as TDD is evaluated in different programming environments) and to increase the internal validity of the results (as the performance of participants forced to use an unfamiliar programming environment could be affected).  

In all experiments, participants had two and a half hours to develop each experimental task. One experimental task corresponds to one experimental session. Participants work individually on each experimental task. To avoid a possible effect of language, the experiment is run in English only when the participants are fluent in it (i.e., companies where developers use English in their daily work, university courses taught in English, or university courses where a certain level of English is required). Otherwise, it is run in the participants' mother tongue (this was the case for University 3 and Company 4).

Table \ref{tab:experimental_design} summarizes the characteristics of the experiments in our family. To ease the visualization of the data in Table \ref{tab:experimental_design}, we grouped the experiments by experimental design.

\subsection{Ethics}
\label{ethics}

All participants are aware that they are participating in an experiment measuring the quality of their solutions. Participation is voluntary in all experiments except University 6 and University 7. In all companies and University 1, participants are enrolled in a training course about TDD. Students in University 2, University 3, University 4, University 5 and University 8 participate in a voluntary exercise. In University 6 and University 7, the experiment is aligned with the teaching goals of the course.

In University 1, University 4 and University 8, the participation in the experiment automatically implied giving consent for using the participants’ data for analysis. In University 2, University 3, University 5, university 6 and University 7 a written consent was signed. Note that the GitHub repository contains only the code of those participants who gave us consent.

We assured participants before running the experiments that their data would be anonymized so that they could not be identified by any means. According to the guide on good data protection practice in research\footnote{https://www.eui.eu/documents/servicesadmin/deanofstudies/researchethics/guide-data-protection-research.pdf}, an effective anonymization solution prevents all parties from singling out an individual in a dataset. The suggested procedures for anonymization are: 1) use random and unpredictable pseudonyms, 2) make sure the number of pseudonyms possible is so large that the same pseudonym is never randomly selected twice. Our anonymization procedure consists of two steps:

\begin{enumerate}

	\item We provided our contact person in each company with the format of the IDs to be generated. We explained them that we neither want to know the participants’ names, nor the link between the IDs and the people. Participants were requested to use their IDs during the whole experiment. 
	
	\item We applied the (irreversible) cryptographic hash function MD5 to the IDs generated by the companies.

\end{enumerate}

This two-step procedure guarantees that neither the general public, nor the companies, nor the researchers know who the participants are. Therefore, the link between an individual subject and a performance measure in our dataset is completely lost. For the case of the experiments run with students, it is possible that the researchers who conducted that specific experiment know who they are, but neither the rest of the researchers involved in this project, nor the general public.

\subsection{Participants}
\label{participants_description}

Participants were handed a survey some days before the experiments took place. The survey contained a series of self-assessment questions that asked the participants about their experience with programming, unit testing, the programming language, and the testing tool to be used during the experiment. We followed an approach similar to the one suggested by Falessi et al. \cite{falessi2018empirical} to measure the real participant experience. In particular, we designed each question so it could be answered on an ordinal scale (inexperienced if the participants had less than 2 years of experience, novice if the participants had between 2 to 5 years of experience, intermediate if the participants had between 5 to 10 years of experience, and expert if the participants had more than 10 years of experience). 

After using this questionnaire in several experiments (University 1, 4 and 8, and Company 1 and 2), we realized that the periods of time that we attached to each level of experience (experience of less than 2 years, 2 years to 5 years, 5 to 10 years, and greater than 10 years) were not suitable for distinguishing participants with completely different experience. Indeed, according to this scale, most of the participants to fell into identical experience levels (as there were hardly any participants with more than 5 years of unit testing or testing tool experience). In turn, these scales were not fine-grained enough to distinguish between differently experienced participants, even if they used unit testing and testing tools as frequently as each other.

In view of this, and as no consensus seems to have been reached yet on how best to measure developer experience in SE \cite{bergersen2014construction,falessi2018empirical,feigenspan2012measuring}, we decided to remove the time periods that we had allotted to each experience level. In fact, we adopted a similar approach to the one suggested by Feigenspan et al. \cite{feigenspan2012measuring} to measure participant experience in the rest of experiments that we ran (University 2, 3, 5, 6 and 7, and Company 3 and 4): with self-assessment questions on an ordinal scale (inexperienced, novice, intermediate and experts) without time-associated experience levels. Although this decision helped us to better describe participants (as it provided for a greater variability of participant responses---see below), this change of scales had an impact on the possibility of assessing the effect of participant experience in the joint meta-analysis of results. In particular, joint meta-analyses run the risk of mixing apples and oranges, as participant experience was measured differently across the experiments. We discuss the trade-offs of this decision in the threats to validity section. 

\begin{figure}[h!]
    \caption{Profile plot: participant experience with first questionnaire.}
    \label{profile_1}
    \centering
    \includegraphics[width=9cm,keepaspectratio]{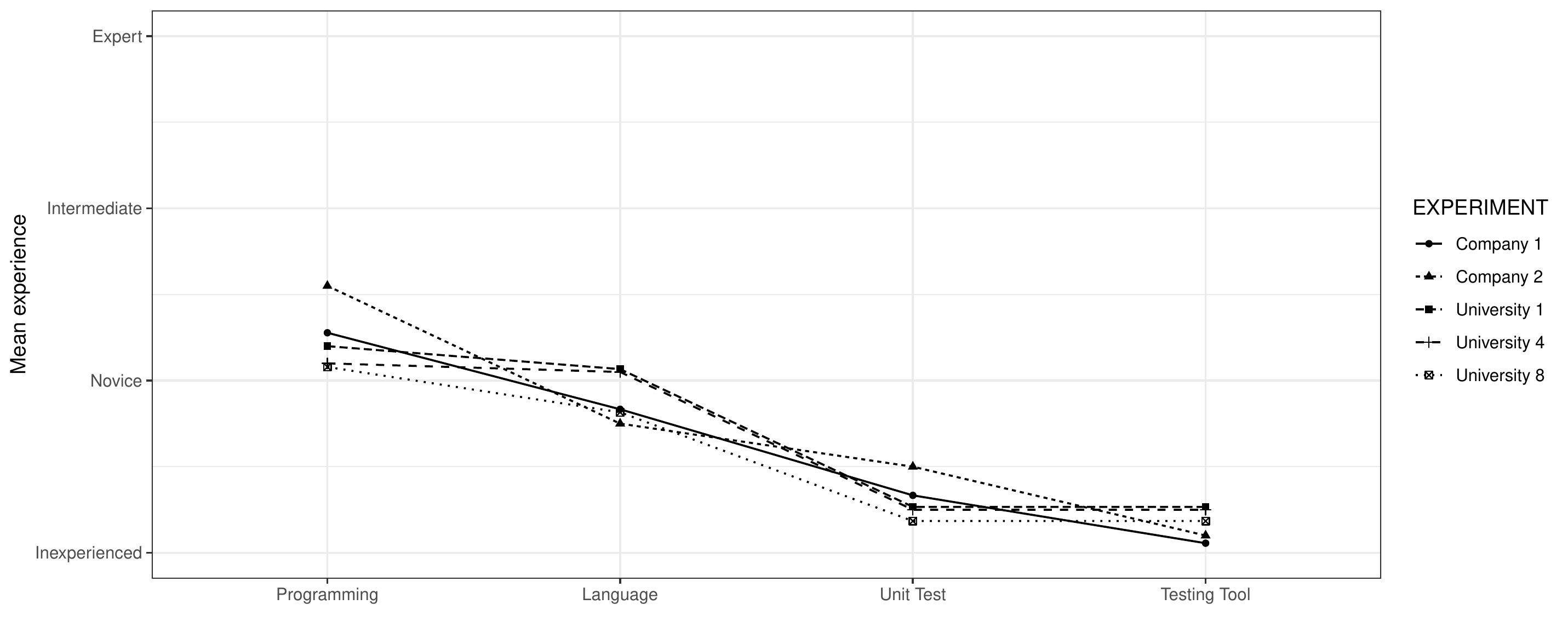}
\end{figure}

\begin{figure}[h!]
    \caption{Profile plot: participant experience with second questionnaire.}
    \label{profile_2}
    \centering
    \includegraphics[width=9cm,keepaspectratio]{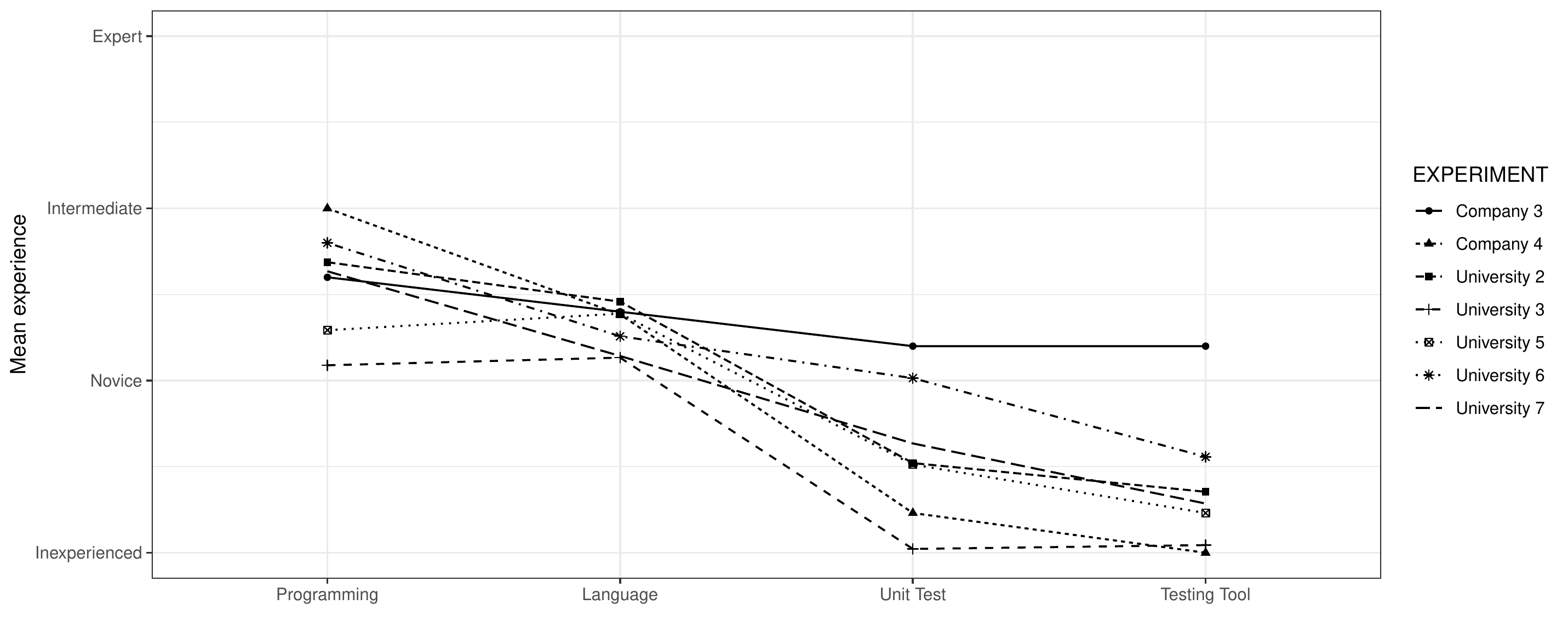}
\end{figure}

Figure \ref{profile_1} shows the profile plot illustrating the mean participant experience in the experiments using the first type of questionnaire. Figure \ref{profile_2} shows the profile plot for participant experience using the second type of questionnaire. For simplicity's sake, we consider here, for merely descriptive purposes, that inexperienced, novice, intermediate and expert correspond to 1, 2, 3 and 4, respectively, and that experience is measured on a continuous rather than an ordinal scale. This approach is typically followed in other disciplines \cite{norman2010likert}. Participant TDD and ITL experience has not been included in Figures \ref{profile_1} and \ref{profile_2} because participants had no previous experience with TDD or ITL.  

We find in both Figure \ref{profile_1} and Figure \ref{profile_2} that participants claimed to have more programming than programming language experience during the experiment. Also, participants claimed to be more experienced with programming or the programming language than with unit testing or the testing tool used during the experiment. Besides, while participant experience across the experiments in Figure \ref{profile_1} is \textit{jammed} together (as, remember, most of the participants using the first type of questionnaire reported having similar levels of experience because experience was associated with years of practice), the spread of participant experience is less pronounced than in Figure \ref{profile_2}. Finally, while the participants with the greatest programming experience are based at Company 2 in Figure \ref{profile_1} and  at Company 4 in Figure \ref{profile_2} (i.e., professionals in both cases), there is a tendency for professionals to consider themselves less skilled than students in some cases (e.g., notice that the programming language experience line at University 1 is above that of Companies 1 and 2 in Figure \ref{profile_1} or that the line for all experience variables, except programming experience, at Company 4 is lower than in most experiments with students in Figure \ref{profile_2}). This must be taken into consideration for exploratory analysis purposes, as it may have an impact on their results.

None of the professionals claimed to have any prior experience with either ITL or TDD. As the experiments were embedded within training courses on TDD and ITL, the professionals would have skipped the course---and the experiment altogether---if they had had any previous experience in these development approaches. Considering that course attendance was optional, professionals already familiar with TDD and ITL may have not seen any benefit in taking time out to attend the course and take part in the experiment.

As a summary, according to the results of the questionnaires and our perceptions while running the experiments, the sample of participants in our family had disparate programming, programming language, unit testing and testing tool experience. Additionally, and as all the experiments were embedded within training courses, all the participants were absolute TDD novices. 

\subsection{Analysis Approach}
\label{analysis_approach}

As regards \textbf{RQ1}, we first provide the descriptive statistics (sample sizes, means, standard deviations and medians \cite{field2013discovering}) of the quality scores achieved with TDD and ITL in each experiment. With the aim of easing the understanding of the data, we then round out the descriptive statistics with a profile plot showing the mean quality scores achieved with ITL and TDD in each experiment. Afterwards, we report the joint result for all the experiments together output by a two-stage individual participant data (IPD) meta-analysis \cite{riley2010meta}. In particular, we first analyze the data of each between-subjects experiment with an independent $t$-test \cite{field2013discovering}, and the data of each within-subjects experiment with a dependent $t$-test \cite{field2013discovering}. Then, we combine the estimates provided by the $t$-tests (i.e., the difference between the means of the TDD and ITL groups), and their respective standard errors by means of random-effects (RE) model meta-analysis \cite{borenstein2011introduction}. We adopted this approach as it allows us to:

\begin{itemize}
	
	 \item Combine the results of experiments with different experimental designs into a joint conclusion.
	 
	 \item Transparently weight each experiment in the joint result according to the standard error of its estimate.
	 
	 \item Provide visual summaries of results (i.e., forest plots \cite{borenstein2011introduction}).
	 
	 \item Incorporate heterogeneity by simply fitting a random-effects model instead of a fixed-effects model \cite{borenstein2011introduction}.
	 
	 \item Assess heterogeneity with relatively straightforward means: the $Q$-test or the $I^2$ statistic. $Q$ tests the null hypothesis that all studies share a common effect size. $I^2$ determines what proportion of the observed variance is real (if it is near zero, almost all the observed variance is spurious, and there is nothing to explain; if it is large, it would make sense to speculate about reasons for, and try to explain, the variance). $I^2$ is interpreted as: 25\% low, 50\% medium and 75\% high \cite{borenstein2011introduction}.  
	
\end{itemize}

Regarding \textbf{RQ2}, we consider the data of MR and BSK only. The participants in the experiments of our family were seldom set the SDK, MP and SS tasks. Therefore, we did not have sufficient data to get accurate results (we acknowledge the limitations of this decision in the threats to validity section). We analyze the data by means of a one-stage IPD meta-analysis \cite{riley2010meta}; in particular, we fit linear mixed models \cite{brown2014applied}. As we did in the previous meta-analysis, we allow for heterogeneity of results by considering the treatment as a random effect across the experiments. To assess the influence of task on quality we fit a linear mixed model with the main effects of treatment and task and their interaction. To ease the understanding of the results, we provide the respective ANOVA tables \cite{field2013discovering} and marginal means \cite{field2013discovering} for each analysis. We use a one-stage IPD meta-analysis rather than a two-stage IPD meta-analysis (as we did before) because \cite{brown2014applied,riley2010meta}:

\begin{itemize}

	\item It is useful for assessing the effect of multiple variables on results at once (e.g., the main effects for treatment and task, and their interaction).
	
	\item It outputs personalized contrasts for interpreting the results (e.g., what is the expected quality with either TDD or ITL for the average number of tests developed by the participants?).
	
	\item It accommodates missing data so that participants with missing scores can still contribute to the joint result (e.g., by fitting linear mixed models \cite{brown2014applied}). 

\end{itemize}

We analyze \textbf{RQ3} by means of RE model sub-group meta-analysis \cite{borenstein2011introduction}. We round out each of the sub-group meta-analyses that we perform with an IPD one-stage linear mixed model with interaction terms \cite{brown2014applied}. We use linear mixed models to provide marginal means and, thus, ease the interpretation of the results. In the following, for each variable that we study (i.e., participant type, programming environment, etc.), we report the results of the sub-group meta-analysis that we performed, its corresponding forest plot, and the marginal means that we calculated from the linear mixed model. We adopt this approach as it allows us to: 

\begin{itemize}

	\item Visually convey the results by means of forest plots \cite{borenstein2011introduction}.
	
	\item Relatively straightforwardly interpret the heterogeneity of the results (e.g., using the $Q$-test or the $I^2$ statistic \cite{borenstein2011introduction}). 
	
\end{itemize}

Finally, we analyze \textbf{RQ4} by means of a one-stage IPD meta-analysis, as we did for RQ2, for the same reasons. As above, we account for heterogeneity of results by considering the treatment as a random effect across the experiments. To assess the influence of participant characteristics on quality, we fit four different linear mixed models (one for each of the four characteristics) as specified in \cite{fisher2011critical},  with the main treatment effects and the respective characteristic, their interaction, and the interaction between the treatment and the mean experience in each experiment. To ease the understanding of the results, we report the ANOVA p-values, marginal means, and profile plots for the interaction between treatment and participant characteristic. Separate analyses are performed for students and professionals, and for the two different types of questionnaires used in the experiments, as, according to Section \ref{participants_description}, they could be influencing the results.

\section{RQ1: ITL vs. TDD}
\label{rq1}

Throughout this section, we answer RQ1: Do TDD and ITL perform similarly in terms of quality? To do so, we present the descriptive statistics of the data in Section \ref{descriptive_statistics_main}, the results of the joint analysis in Section \ref{analysis_main}, and discuss our findings, which we frame within the results of the identified primary studies on TDD, in Section \ref{findings_main}.

\subsection{Descriptive Statistics}
\label{descriptive_statistics_main}

Table \ref{stargazer_summary} shows the sample sizes (N), means, standard deviations (SD) and medians of the quality scores achieved with ITL and TDD in all the experiments within our family (in the same order as in Table \ref{tab:experimental_design}). As we can see, the mean quality achieved with ITL goes from as high as $M=68.28$ at Company 1 to as low as $M=22.84$ at Company 4, while the means for TDD go from as high as $M=67.64$ at Company 3 to as low as $M=10.68$ at Company 4. 

\begin{table}[!h] \centering 
  \caption{Summary statistics for quality by experiment and development approach.} 
  \label{stargazer_summary} 
\begin{tabular}{llrrrr} \hline \hline 
\textbf{Experiment} & \textbf{Treatment} & \textbf{N} & \textbf{Mean} & \textbf{SD} & \textbf{Median} \\ \hline 
\multirow{2}{*}{University 1} & ITL & $16$ & $32.12$ & $29.94$ & $24.11$ \\ 
& TDD & $14$ & $31.85$ & $21.79$ & $24.56$ \\ \hline
\multirow{2}{*}{Company 1} & ITL & $16$ & $68.28$ & $34.27$ & $85.71$ \\ 
& TDD & $17$ & $42.50$ & $39.32$ & $42.70$ \\ 
\multirow{2}{*}{Company 2} & ITL & $20$ & $41.61$ & $31.66$ & $50.75$ \\ 
 & TDD & $20$ & $38.28$ & $22.45$ & $33.33$ \\ 
\multirow{2}{*}{University 2} & ITL & $44$ & $65.34$ & $30.08$ & $76.56$ \\ 
 & TDD & $43$ & $36.24$ & $28.80$ & $39.33$ \\ 
\multirow{2}{*}{University 3} & ITL & $49$ & $26.52$ & $18.11$ & $23.44$ \\ 
& TDD & $52$ & $32.36$ & $25.05$ & $33.26$ \\ \hline
\multirow{2}{*}{Company 3} & ITL & $8$ & $50.42$ & $32.76$ & $46.36$ \\ 
& TDD & $6$ & $67.64$ & $26.24$ & $70.77$ \\ 
\multirow{2}{*}{Company 4} & ITL & $12$ & $22.84$ & $20.57$ & $17.12$ \\ 
& TDD & $11$ & $10.69$ & $14.75$ & $7.86$ \\ \hline 
\multirow{2}{*}{University 4} & ITL & $20$ & $38.20$ & $31.16$ & $39.43$ \\ 
& TDD & $20$ & $41.42$ & $35.58$ & $35.79$ \\ \hline
\multirow{2}{*}{University 5} & ITL & $16$ & $43.86$ & $23.85$ & $52.68$ \\ 
 & TDD & $18$ & $39.35$ & $25.03$ & $29.46$ \\
\multirow{2}{*}{University 6} & ITL & $33$ & $37.12$ & $30.08$ & $29.69$ \\ 
& TDD & $38$ & $39.52$ & $27.48$ & $29.69$ \\ 
\multirow{2}{*}{University 7} & ITL & $35$ & $44.96$ & $28.97$ & $23.44$ \\ 
& TDD & $28$ & $34.26$ & $20.53$ & $23.44$ \\ 
\multirow{2}{*}{University 8} & ITL & $20$ & $37.42$ & $39.31$ & $22.22$ \\ 
& TDD & $21$ & $29.24$ & $36.66$ & $5.62$ \\ \hline
\end{tabular} 
\end{table} 

\begin{figure}[h!]
    \caption{Profile plot: ITL and TDD mean quality.}
    \label{profile_plot}
    \centering
    \includegraphics[width=9cm,keepaspectratio]{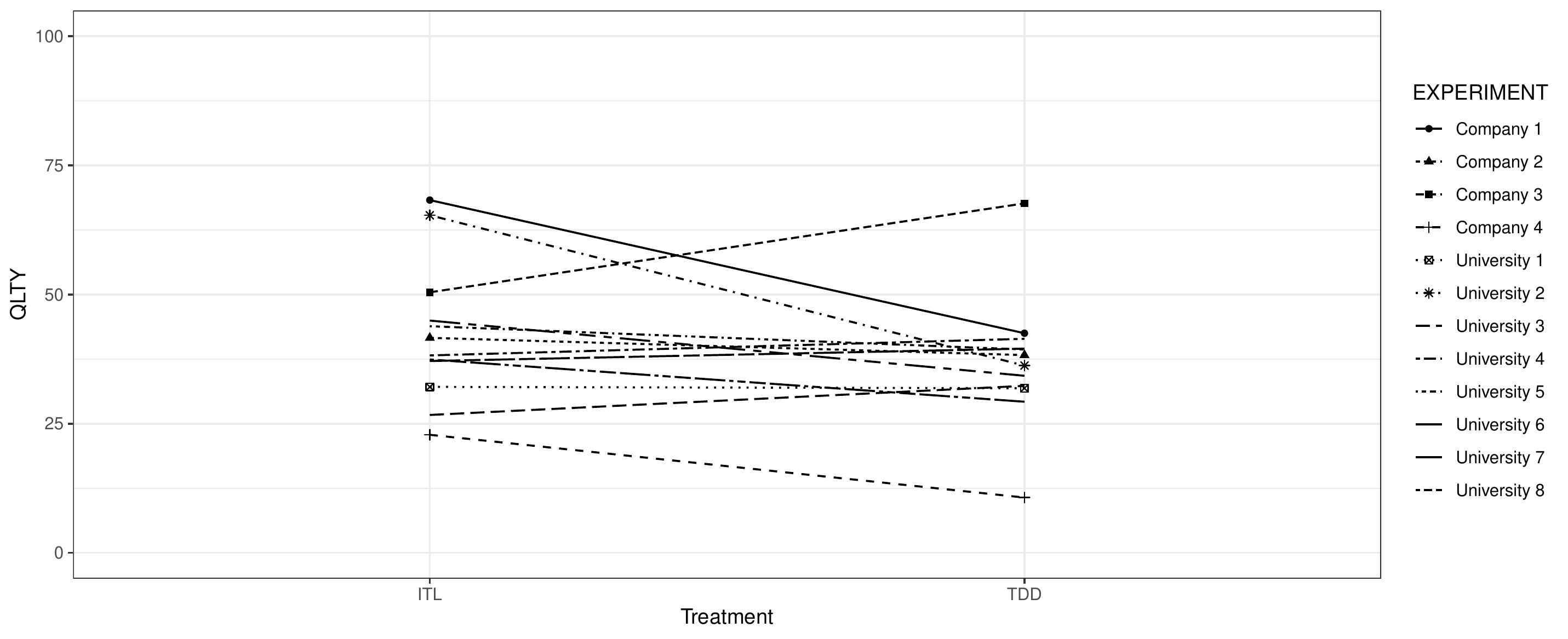}
\end{figure}

Figure \ref{profile_plot} shows the profile plot for the mean quality scores achieved with ITL and TDD in all the experiments\footnote{We analyzed the data with t-tests. Therefore, the mean difference (i.e., the slope of the line) provides useful information for evaluating experiment results.}. As we can see, the means for TDD and ITL are clustered between 50 and 25 in most experiments. Additionally, the slope of the lines in these experiments (i.e., the difference between the mean performance achieved with TDD and ITL) is more or less flat. In other words, TDD and ITL appear to perform similarly in these experiments. However, some experiments have relatively larger slopes (i.e., Universities 3 and 5 and Companies 1 and 3), where the differences between TDD and ITL are more noticeable. 

\subsection{Analysis for ITL vs. TDD}
\label{analysis_main}

\begin{figure*}[h!]
    \caption{Forest plot: TDD vs. ITL mean effects.}
    \label{ad_mean}
    \centering
    \includegraphics[width=12cm,keepaspectratio]{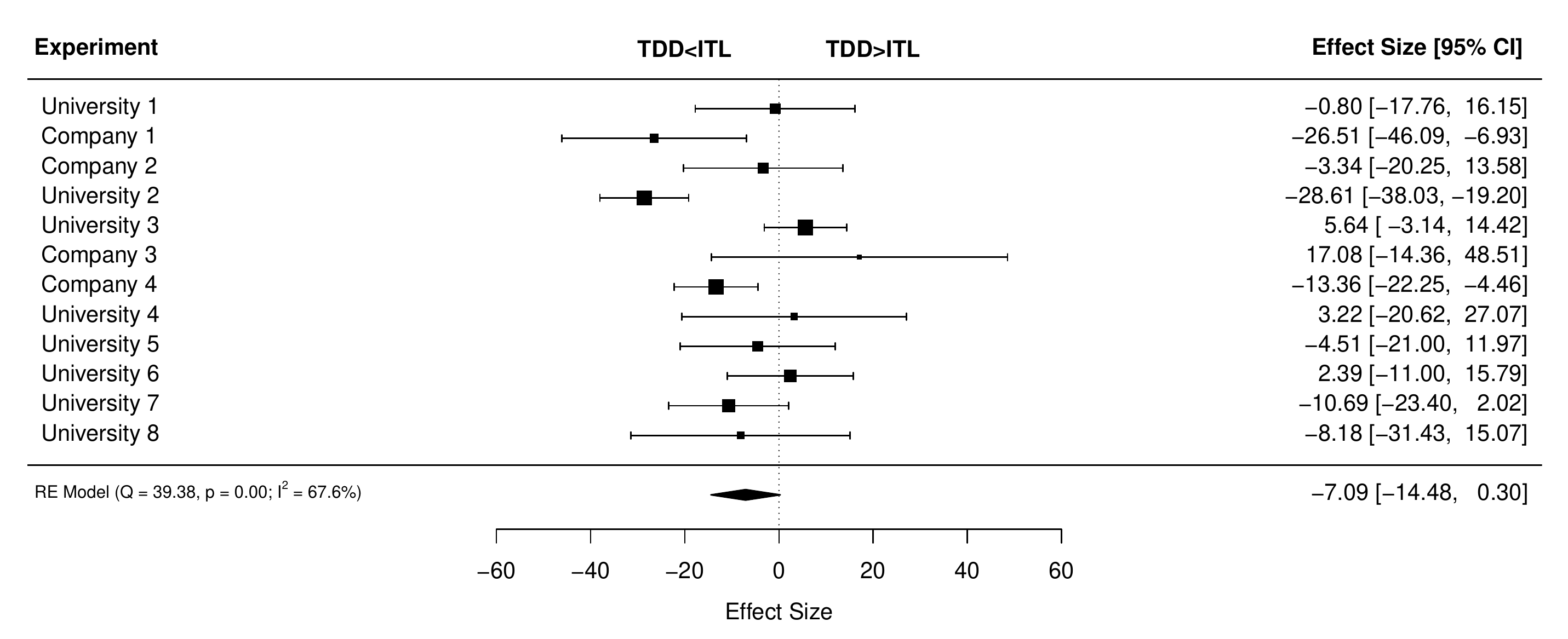}
\end{figure*} 

Figure \ref{ad_mean} shows the forest plot for the RE model meta-analysis that we performed. It shows that ITL achieves higher quality than TDD in most experiments (as the estimate for these experiments is on the left-hand side of the figure). In only three experiments (i.e., Companies 1 and 4, and University 2), however, is the difference in performance between ITL and TDD statistically significant (statistical significance holds when the 95\% confidence intervals---95\% CI---of the estimate do not cross 0). When pooling together all the estimates by means of meta-analysis (see the black diamond at the bottom), ITL achieves higher quality than TDD ($M=-7.06$, 95\% \textit{CI}=(-14.47, 0.35)). However, this difference in performance is not statistically significant (as the 95\% CI crosses 0), and \textit{small}---considering that $M=-7.06$ is a small drop in performance compared to the full range of quality scores that can be achieved (i.e., 100\% to 0\%). Additionally, there is a \textit{medium} heterogeneity of results ($I^2$=67.8\%, $Q$-statistic=39.70, $p$-value=$<$0.001) according to commonly used rules of thumb \cite{borenstein2011introduction}.  \\

\noindent\fbox{
  \parbox{11.5cm}{
    \textbf{Key findings}
    \begin{itemize}
        \item{The joint result shows that TDD novices coding toy tasks achieve slightly better quality scores with ITL than with TDD.}
        \item{The results vary largely across the individual experiments.}
    \end{itemize}
    }
}

\subsection{Discussion}
\label{findings_main}

The participants achieve low quality scores overall, regardless of the treatment, despite the fact that they are coding toy tasks. However, the standard deviation is also large. We hypothesize that the low experience of the participants could explain their low quality scores. More precisely, the testing experience might have played a critical role. Participants must write test cases regardless of the treatment being applied. If their unit testing experience is low, the overhead caused by writing the test cases could have reduced the time necessary to produce high quality code.

ITL quality is \textit{slightly} higher than for TDD in our family of experiments. However, this difference is not statistically significant. Nonetheless, as the 95\% CI of the joint result in our family (i.e., the black diamond at the bottom of the forest-plot) is very narrow---at least compared to the 95\% CI of the individual experiments---running another experiment and adding its results to the outcomes of our family is unlikely to make much difference the joint result much. In particular, as we already conducted a joint meta-analysis of the data of 12 experiments in our family, our experiments would have a relatively greater weight than an isolated experiment in the joint result of any prospective meta-analysis (unless, of course, a huge experiment was added to our family \cite{borenstein2011introduction}). Consequently, the joint results would be similar to the findings for our family.

Our joint results (i.e., the control approach achieved slightly higher quality than TDD) are consistent with findings reported in other experiments on TDD where MSc students developed toy tasks in experiments spanning days or weeks \citeP{mueller2002experiment,zielinski2005preliminary}. Additionally, the results achieved at Company 3 (i.e., quality for TDD is slightly higher than for the control approach) are also consistent with the outcomes of other experiments with professionals published in the literature (i.e., \citeP{george2004structured}). However, Company 3 results should be interpreted with caution, at least in view of the wide 95\% CI that materialized (meaning that many other results may be compatible with the data). The extreme results for Company 3 may be due to small sample size, where natural variability is highly likely to be behind the outliers.

When comparing the results of our family with the findings of the experiments on TDD published so far, we conclude that there does not appear to be much difference for TDD novices between TDD and the control approach (since the difference in our family is small, as are the both positive and negative differences in most of the published experiments). Thus, to date, at least, \textit{TDD novices seem to perform similarly or slightly better with the control approach than with TDD in experiments}. This contrasts with the results usually reported in case studies, where vast improvements are usually achieved with TDD (ranging from 22\% \citeP{edwards2003using} to 267\% \citeP{ynchausti2001integrating}). We hypothesize that at least the following reasons may be influencing this disparity.

First, we think that the duration of the studies may be influencing the results. In particular, as experiments usually span only a few days or weeks and case studies usually span months or years, case study participants benefit more from TDD as a result of a longer exposure to the approach. In other words, TDD may only show its benefits after it has been internalized by developers, and this process may take more than a few days or weeks. To check this hypothesis, we propose to run further experiments with TDD novices that have been trained in, and have used, TDD for several months (or at least weeks) before the experiment takes place. We also propose to run experiments with TDD experts vs. control approach (e.g., waterfall model or ITL) experts to emulate the case study situation---where participants are already knowledgeable about both development approaches.

Second, we single out the unit of analysis (i.e., solo, pair programming, or teams) as a possible reason behind the different results reported in the literature. According to the results that we observed in the literature at least, the studies that measured quality for the whole development team (i.e., measured the quality of a system after it had been developed by a team of developers) achieved more optimistic results for TDD than others that measured quality just for solo or pair programmers. To check this hypothesis, we propose to run further experiments to compare the performance of solo programmers, pair programmers and teams with TDD again to emulate the case study situation.

Third, we think that the mixed profile of the participants in case studies (different participants have different experience levels) is another possible reason for the difference in the results that we observed between the literature and our family of experiments.

Fourth, the real benefits of TDD could be due to short development cycles rather than test-first development. This has been observed by Fucci et al. \cite{fucci2017dissection} in one of the experiments cited in this paper. This would explain why TDD performs better when the waterfall model, instead of ITL, is used as the control.

Finally, we should underscore the fact that, despite our attempts to keep the experimental configurations of the experiments as alike as possible (although we were obliged to adapt their programming environments and change their experimental designs at times), there was still a noticeable heterogeneity of outcomes within our family. This indicates that there is a slight inconsistency across the results of the different studies. Consequently, moderator variables that could be affecting the outcomes need to be studied, as heterogeneity suggests that the characteristics of the experiments may be impacting their results. In the following, we try to identify the extent to which the different characteristics of the experiments are causing the detected heterogeneity of results. 

We do not think that companies developing different types of software perform differently. While Company 1 and Company 3 develop betting software for online gaming, Company 2 develops electronics software, and Company 4 develops software for supplies management (water, internet, telephone and television). If this were the case, Company 1 and Company 3 should show consistent results. But according to Figure~\ref{ad_mean}, Company 1 and Company 3 have very different results. However, this could be explained also by the low sample size of Company 3. On the other hand, Company 1, Company 2 and Company 4 seem to show consistent results.

Even though we do not focus on the length and rhythm of the development cycles with TDD (see \cite{fucci2017dissection} for an extended discussion on this point), our results appear to be consistent with the observations made by Karac and Turhan \cite{karac2018we}. In particular, both TDD and ITL perform similarly within our family of experiments. Besides, working on small, well-defined tasks (such as BSK, rather than MR) also seems to translate into higher external quality.

\section{RQ2: Task}
\label{rq2}

Throughout this section we answer RQ2: To what extent is quality affected by the task under development? We think that this is a relevant issue. Although MR and BSK are of similar complexity, their specifications do not have the same granularity (BSK is fine-grained, while MR is coarse-grained). This could benefit either (ITL or TDD) or both of the approaches in terms of quality. In order to answer this research question, we present the descriptive statistics and the analysis for the task in Section~\ref{analysis_task}. Then, we discuss and compare our findings with the outcomes published in the literature in Section~\ref{findings_task_test}.

\subsection{Analysis for Task}
\label{analysis_task}

Table \ref{task_descriptive} shows the sample sizes (N), means, standard deviations (SD) and medians of the quality scores achieved with each task in all the experiments within our family (experiments are organized in the same order as in Table \ref{tab:experimental_design}). The mean quality achieved with BSK goes from as high as $M=65.34$ at University 2 to as low as $M=23.81$ at Company 4. The means for MR go from as high as $M=67.11$ at Company 3 to as low as $M=7.46$ at University 8.

\begin{table}[!h] \centering 
  \caption{Summary statistics for quality by experiment and task.}
  \label{task_descriptive} 
\begin{tabular}{cccccc} \\ \hline \hline 
\textbf{Experiment} & \textbf{Task} & \textbf{N} & \textbf{Mean} & \textbf{SD} & \textbf{Median} \\ \hline 
\multirow{4}{*}{University 1} & MR & $8$ & $10.96$ & $17.31$ & $2.81$ \\ 
& BSK & $8$ & $50.22$ & $24.94$ & $43.75$ \\ 
& MP & $7$ & $25.44$ & $13.21$ & $24.39$ \\ 
& SDK & $7$ & $41.76$ & $28.73$ & $23.08$ \\ \hline
\multirow{2}{*}{Company 1} & MR & $16$ & $44.24$ & $39.34$ & $45.51$ \\ 
& BSK & $17$ & $65.13$ & $36.25$ & $83.93$ \\ 
\multirow{2}{*}{Company 2} & MR & $20$ & $53.53$ & $25.55$ & $63.73$ \\ 
& BSK & $20$ & $26.36$ & $21.71$ & $22.14$ \\ 
\multirow{2}{*}{University 2} & MR & $43$ & $36.24$ & $28.80$ & $39.33$ \\ 
& BSK & $44$ & $65.34$ & $30.08$ & $76.56$ \\ 
\multirow{2}{*}{University 3} & MR & $48$ & $17.37$ & $16.27$ & $9.55$ \\ 
& BSK & $53$ & $40.54$ & $20.86$ & $32.81$ \\ \hline
\multirow{3}{*}{Company 3} & MR & $6$ & $67.11$ & $40.02$ & $81.46$ \\ 
& BSK & $4$ & $58.48$ & $26.10$ & $59.82$ \\ 
& SS & $4$ & $43.16$ & $13.45$ & $46.36$ \\ 
\multirow{3}{*}{Company 4} & MR & $8$ & $12.64$ & $14.89$ & $8.43$ \\ 
& BSK & $7$ & $23.81$ & $26.64$ & $12.82$ \\ 
& SS & $8$ & $15.48$ & $14.00$ & $10.71$ \\ \hline
\multirow{2}{*}{University 4} & MR & $20$ & $17.74$ & $24.87$ & $7.30$ \\ 
& BSK & $20$ & $61.89$ & $24.46$ & $70.18$ \\ \hline
University 5 & BSK & $34$ & $41.47$ & $24.22$ & $45.54$ \\ 
University 6 & BSK & $71$ & $38.40$ & $28.54$ & $29.69$ \\ 
University 7 & BSK & $63$ & $40.20$ & $25.93$ & $23.44$ \\ 
\multirow{2}{*}{University 8} & MR & $22$ & $7.46$ & $20.54$ & $0.56$ \\ 
& BSK & $19$ & $63.07$ & $30.30$ & $76.19$ \\  \hline 
\end{tabular} 
\end{table}

Table \ref{anova_task} shows the ANOVA table corresponding to the analysis that we undertook to assess the influence of task and treatment on quality. As we can see, the task under development has a statistically significant effect on quality. This suggests that \textit{the task being developed influences quality consistently} (i.e., either increasing or decreasing the quality scores achieved with \textit{both} development approaches), meaning that participants achieve higher quality in some tasks compared to others. Besides, as the interaction between the task and the treatment is not statistically significant, this means that the task under development is not responsible for the difference in performance between TDD and ITL---as, after all, the task has a similar impact (either increasing or decreasing) on the quality scores of both development approaches.

\begin{table}[!h] \centering 
  \caption{ANOVA table for quality: task by treatment.} 
  \label{anova_task} 
\begin{tabular}{lccc} \hline \hline 
\textbf{Factor} & \textbf{Chisq} & \textbf{Df} & \textbf{$p$-value}\\ \hline 
Treatment & 1.148 & 1 & 0.28 \\
Task & 60.05 & 1 & $<$.001 \\
Treatment:Task & 0.08 & 1 & 0.78 \\ \hline
\end{tabular} 
\end{table}

We complement the results of the ANOVA table with the marginal means of the interaction (i.e., the average quality scores achieved with each combination of task and treatment) in Table~\ref{marginal_means_task}.

\begin{table}[!h] \centering 
  \caption{Marginal means for quality: task by treatment.} 
  \label{marginal_means_task} 
\begin{tabular}{llcc} \hline \hline 
\textbf{Task} & \textbf{Treatment} & \textbf{Estimate} & \textbf{95\% CI}\\ \hline 
\multirow{2}{*}{MR} & ITL & 29.36 & (17.65, 41.06) \\
 & TDD & 24.95 & (16.95, 32.97)  \\ \hline
 \multirow{2}{*}{BSK}& ITL & 49.22 &  (39.13, 59.30) \\
 & TDD & 46.46 & (38.92, 54.01) \\ \hline
\end{tabular} 
\end{table} 

As we can see in Table~\ref{marginal_means_task}, the quality scores achieved for MR are considerably lower than those achieved for BSK with both development approaches. Besides, the difference in performance between ITL and TDD for both tasks is small---although ITL achieves slightly higher quality than TDD for both MR and BSK. The difference in the level of detail of the specifications may have boosted the performance of the participants when developing BSK (see \cite{karac2019} for further research on this issue). \\

\noindent\fbox{
  \parbox{11.5cm}{
    \textbf{Key findings}
    \begin{itemize}
        \item{Some tasks lead to higher quality scores than others regardless of the development approach applied.}
        \item{The task under development does not appear to be responsible for the \textit{difference} in quality between TDD and ITL.}
    \end{itemize}
    
    }
}

\subsection{Discussion of Task}
\label{findings_task_test}

As we have seen, the task being developed seems to affect consistently---either increasing or decreasing---the quality scores achieved with both TDD and ITL. The reason for this could be tasks having different intrinsic properties that cause some tasks having a greater design complexity/difficulty than others. For example, the MR specification is coarse-grained and BSK is fine-grained. This could explain why BSK gets a better performance. These results are in line with those of Tosun et al. \cite{tosun2020investigating}. On the other hand, there does not appear to be any combination of task and development approach that boosts quality.

This has certain implications. First, as the task being developed seems to affect quality similarly regardless of the development approach applied, the task cannot be blamed for the difference in performance between TDD and the control approach. Extrapolating this finding to the results published in the literature of TDD, we hypothesize that \textit{as long as the same task is developed with TDD and the control approach, the task is unlikely to be behind the heterogeneity of results observed so far}. However, as quality largely depends upon the developed task, if a different task is developed with each development approach (e.g., Task 1-ITL, Task 2-TDD), this may lead to completely different quality scores. This may have happened to us before in one of our experiments on TDD \cite{tosun2017industry}, where, although we thought that BSK and MR were similar, TDD-BSK clearly outperformed ITL-MR. Now that we have decoupled the effect of the task and the treatment in our experiments, we find that the task determines quality to a larger extent than the actual development approach. 

Finding that the task being developed has a noticeable influence on quality has implications in terms of experimental design. In particular, as quality is largely dependent upon the task developed, if different tasks are developed in an experiment, the variability of the quality scores will increase (as either large or small quality scores will be achieved). Thus, if we run a between-subjects experiment, where different tasks are implemented within each treatment, the variability of the quality scores in each treatment will rise, whereas the statistical power will plummet. This will result in wider 95\% CIs (i.e., less precise results) for the experiment \cite{cumming2013understanding,field2013discovering}, the effects of which are shown in the forest plot in Figure \ref{ad_mean}. In particular, notice how, despite Universities 8 and 5 having identical sample sizes (i.e., 41 participants each), the 95\% CI for University 8 (where both BSK and MR were developed with each development approach) is wider than for University 5 (where only BSK was developed with each development approach). In view of this, we recommend either \textit{the use of only one task in between-subjects experiments} or \textit{the substitution of between-subjects experiments for within-subjects experiments if experimenters want to evaluate more than one task}. If within-subjects experiments are run instead of between-subjects experiments, the extra variability introduced by the different tasks should cancel out during data analysis. This is because each participant acts as his or her own baseline in within-subjects experiments, and, thus, the potential correlation between participant scores may counterbalance the extra variability introduced by the different tasks. Additionally, the low mean quality achieved with MR at University 8 could be explained by the fact that University 8 follows a crossover design according to which MR is implemented in the first session and BSK in the second session. The implementation of the most complex/difficult task in the first session could lead to a lower mean. Note that MP and SDK are used only in one experiment (University 1), and SS is used in two experiments only (Company 3 and Company 4).

In summary, there is a need for further studies investigating why task complexity or code visibility impacts quality, along the lines of Tosun et al. \cite{tosun2020investigating}.

\section{RQ3: Experiment-Level Moderators}
\label{rq3}
\begin{figure*}[t!]
    \caption{Forest plot: students vs. professionals}
    \label{subgroup_type_of_participant}
    \centering
    \includegraphics[width=12cm,keepaspectratio]{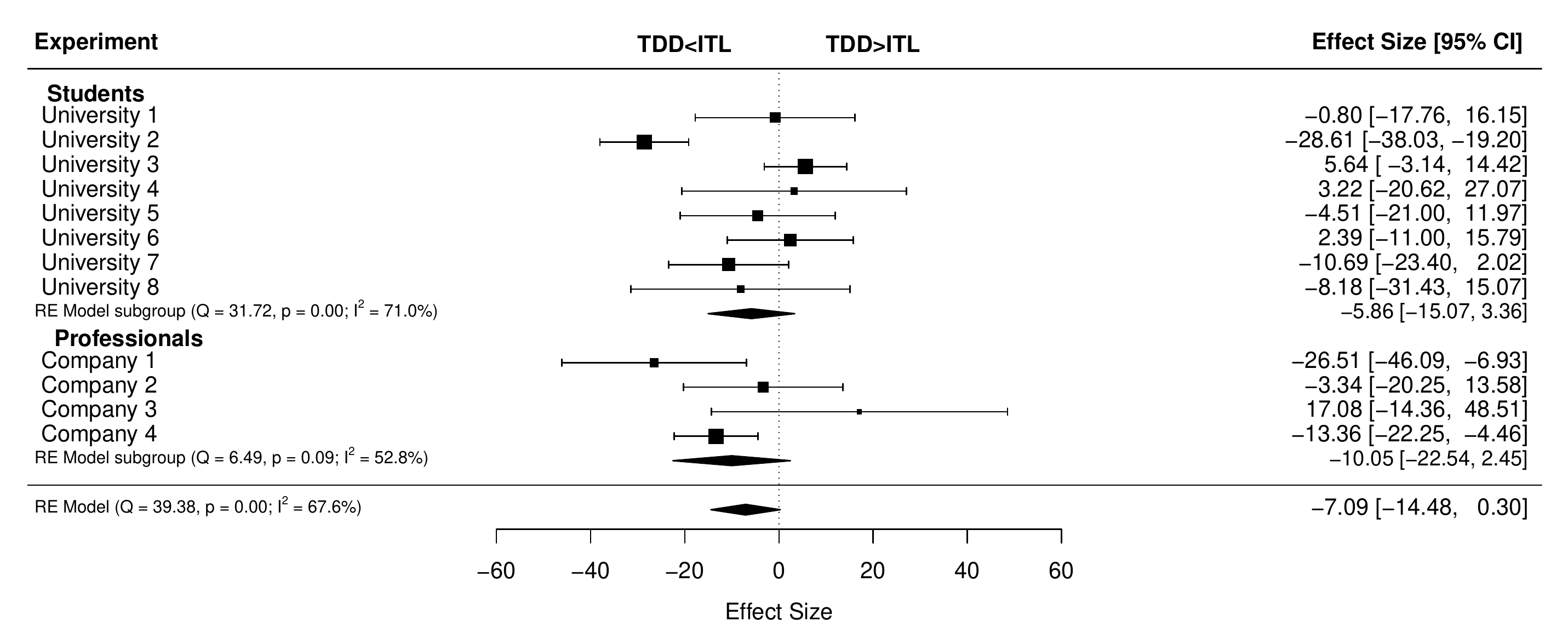}
\end{figure*}

Throughout this section, we answer RQ3: To what extent is quality affected by the characteristics of the experiments: participant type (i.e., professionals vs. students),  programming environment (i.e., Java vs. C++/C\# and related IDEs and testing tools), learning effects from one session to another, or order of application of ITL and TDD? Note that one of the reasons for conducting families of experiments is to hypothesize on experiment-level moderators that may be influencing the results \cite{adriguidelines}. For this purpose, we assess the influence of the participant type, programming environment, learning effects, and the order of application in Sections \ref{analysis_type_of_participants}, \ref{analysis_programming_environment}, \ref{analysis_learning_effects} and \ref{analysis_order_application}, respectively. We then discuss and contextualize our findings in Section \ref{findings_characteristics}.

\subsection{Participant Type}
\label{analysis_type_of_participants}

At the beginning of our article, we hypothesized that participant type (i.e., students vs. professionals) may be influencing the results of the studies published so far on TDD. In view of this, we studied the impact of participant type on the results in our family of experiments. Table \ref{summary_subgroup_type_of_participant} shows a summary of the results of the sub-group meta-analysis that we undertook to assess this point. Figure \ref{subgroup_type_of_participant} shows the corresponding forest plot. As Figure \ref{subgroup_type_of_participant} and Table \ref{summary_subgroup_type_of_participant} show, the magnitude of the joint result for the students sub-group ($M=-5.82$, 95\% \textit{CI}=(-15.06, 3.43)) is smaller than for the professionals sub-group ($M=-10.05$, 95\% \textit{CI}=(-22.54, 2.45)). 

\begin{table}[h!]
\small
\begin{center}
\caption{Sub-group meta-analysis: students vs. professionals.}
\label{summary_subgroup_type_of_participant}
\begin{tabular}{llccc} \hline \hline
\textbf{Group} & \textbf{N} & \textbf{Estimate} & \textbf{95\% CI} & \textbf{$I^2$} \\ \hline
Students & 7 & -5.82 & (-15.06, 3.43) & 71.1\% \\
Professionals & 4 & -10.05 & (-22.54, 2.45) & 52.8\% \\
Difference &  & 4.23 & (-38.23, 14.21)  & - \\ \hline
\end{tabular}
\end{center}
\end{table}

Additionally, as Table \ref{summary_subgroup_type_of_participant} shows, performance for professionals degrades to a larger extent than for students using TDD (the Difference row in Table \ref{summary_subgroup_type_of_participant} indicates that the performance of professionals using TDD drops more than $M=4.23$ points 'on-average' than for students). Thus, in relative terms, \textit{performance by professionals with TDD drops almost twice as much as for students}. We also find that, although participant type does appear to influence results, there is still a noticeable heterogeneity of results in both sub-groups. Thus, other variables rather than participant type may be influencing results. Finally, Table~\ref{marginal_means_students_professionals} shows the estimated quality scores of students and professionals divided by the development approach. 

\begin{table}[!h] \centering 
  \caption{Marginal means for quality: participant type by treatment.} 
  \label{marginal_means_students_professionals} 
\begin{tabular}{llcc} \hline \hline 
\textbf{Treatment} & \textbf{Group} & \textbf{Estimate} & \textbf{95\% CI}\\ \hline 
\multirow{2}{*}{ITL} & Students & 40.47 & (29.19, 51.75) \\
 & Professionals & 49.53 & (31.66, 67.41)  \\ \hline
 \multirow{2}{*}{TDD}& Students & 35.61 &  (30.37, 40.85) \\
 & Professionals & 38.26 & (28.33, 48.18) \\ \hline
\end{tabular} 
\end{table} 

As we can see in Table \ref{marginal_means_students_professionals},  professionals achieve higher quality than students for both TDD and ITL. \\

\noindent\fbox{
  \parbox{11.5cm}{
    \textbf{Key findings}
    \begin{itemize}
        \item{Professionals achieve higher quality than students with both TDD and ITL.}
        \item{However, the drop in performance of professionals applying TDD as compared to ITL almost doubles that of students. }
    \end{itemize} 
    }
}

\subsection{Programming Environment}
\label{analysis_programming_environment}

\begin{figure*}[t!]
    \caption{Forest plot: programming environment.}
    \label{forest_programming_environment}
    \centering
    \includegraphics[width=12cm,keepaspectratio]{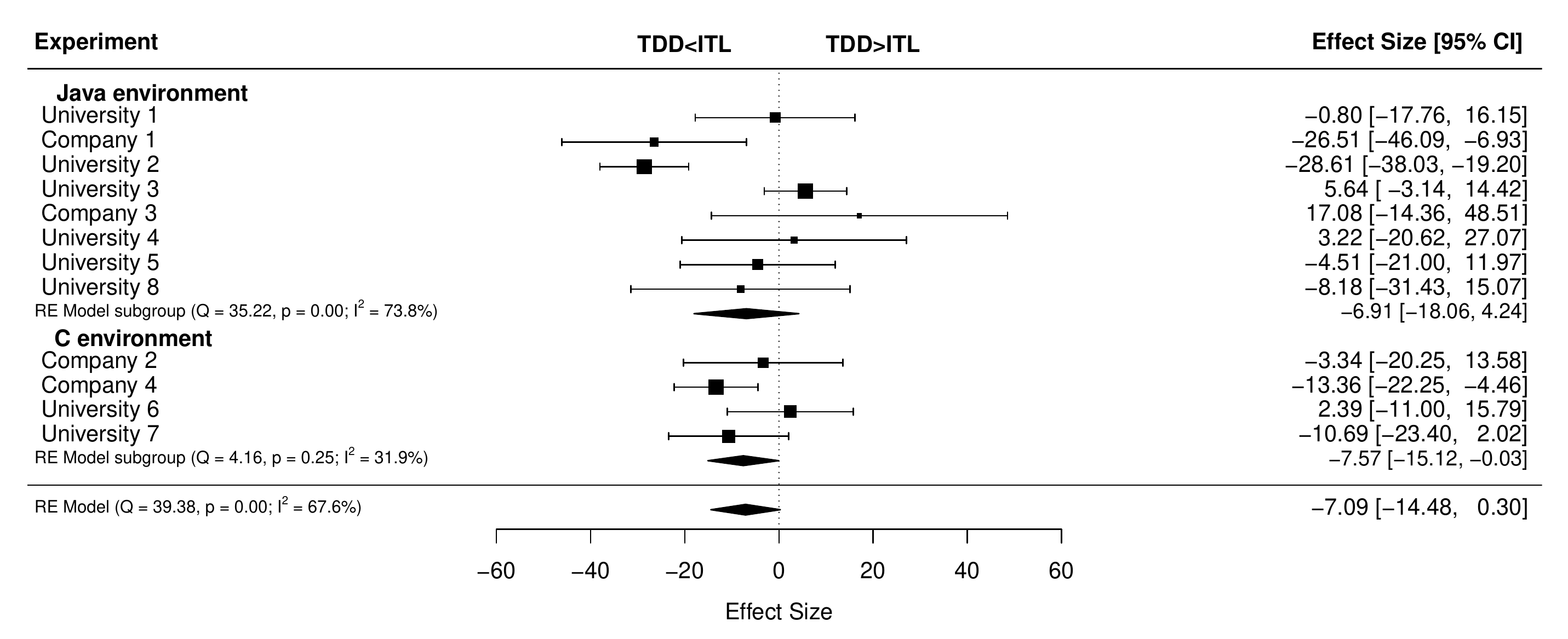}
\end{figure*} 

At the beginning of our article, we hypothesized that the programming environment (i.e., the programming language, IDE, testing tools, etc.) may be behind the heterogeneity of results observed in the literature. Thus, we assessed the extent to which this may be influencing results in our family of experiments. Table \ref{summary_subgroup_programming_environment} summarizes the results of the sub-group meta-analysis that we undertook to assess the influence of the programming environment (i.e., Java/JUnit/Eclipse vs. C++/C\# and related technologies\footnote{In our experiments, the IDEs and testing tools used with C++ and C\# are different. In this study, however, we make the simplification of considering them as being a part of the same group of technologies merely for the purposes of comparison.}) on results. Figure \ref{forest_programming_environment} shows the corresponding forest plot. Table \ref{marginal_means_programming_environment} shows the estimated quality scores for the different programming environments divided by the development approach.

\begin{table}[h!]
\small
\begin{center}
\caption{Sub-group meta-analysis: Java vs. C++/C\#.}
\label{summary_subgroup_programming_environment}
\begin{tabular}{llccc} \hline \hline
\textbf{Group} & \textbf{N} & \textbf{Estimate} & \textbf{95\% CI} & \textbf{$I^2$} \\ \hline
Java & 8 & -6.86 & (-18.05, 4.31) & 74\% \\
C++/C\# & 4 & -7.57 & (-15.11, -0.03) & 31.88\% \\
Difference & - & 0.707 & (-12.78, 14.19)  & - \\ \hline
\end{tabular}
\end{center}
\end{table}

\begin{table}[!h] \centering 
  \caption{Marginal means for quality: programming environment by treatment.}
  \label{marginal_means_programming_environment} 
\begin{tabular}{llcc} \hline \hline 
\textbf{Treatment} & \textbf{Group} & \textbf{Estimate} & \textbf{95\% CI}\\ \hline 
\multirow{2}{*}{ITL} & Java & 43.83 & (32.13, 55.53) \\
 & C\#/C++ & 41.25 & (22.25, 60.26)  \\ \hline
 \multirow{2}{*}{TDD}& Java & 35.77 &  (30.08, 41.47) \\
 & C\#/C++ & 37.55 & (28.82, 46.29) \\ \hline
\end{tabular} 
\end{table} 

As we can see in Table \ref{summary_subgroup_programming_environment}, Figure \ref{forest_programming_environment}, and Table \ref{marginal_means_programming_environment}, ITL achieves higher quality than TDD for both the Java programming environment ($M=-6.86$, 95\% \textit{CI}=(-18.05, 4.31)), and the C++/C\# programming environment ($M=-7.57$, 95\% \textit{CI}=(-15.11, -0.03)). Additionally, the difference between the results achieved with both programming environments is negligible ($M=0.707$, 95\% \textit{CI}=(-12.78, 14.19)). Finally, there is still a considerable amount of heterogeneity in the sub-group formed by the experiments with the Java environment, albeit not as much in the sub-group with C++/C\# technologies. We put this lower $I^2$ statistic in the C++/C\# sub-group down to the small number of experiments and, thus, to the potentially lower accuracy of the $I^2$ statistic \cite{thorlund2012evolution}. 
\\

\noindent\fbox{
  \parbox{11.5cm}{
    \textbf{Key findings} \\
        The programming environment does not appear to affect the difference in performance between TDD and ITL.
    }
}

\subsection{Learning Effects on Treatments}
\label{analysis_learning_effects}

\begin{figure*}[t!]
    \caption{Forest plot: learning effects.}
    \label{forest_learning}
    \centering
    \includegraphics[width=12cm,keepaspectratio]{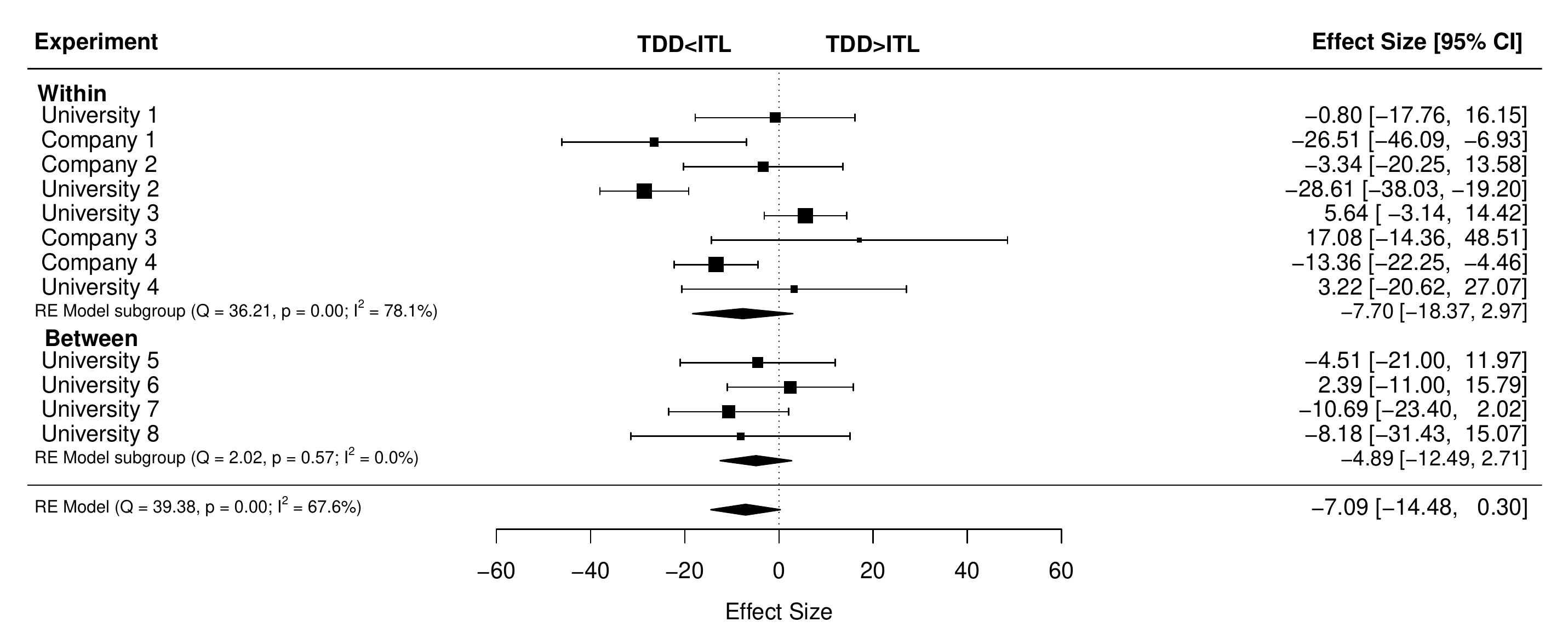}
\end{figure*}

As we acknowledged when discussing the design of our experiments, the participants in within-subjects experiments applied not one but both development approaches. As a result, they could learn something during the first session that could boost their performance in the second session and may influence the results achieved. To assess the extent to which this was the case in our family of experiments, we compared the results of the experiments where the participants applied both development approaches (i.e., both the AB within-subjects experiments and the crossover experiment), with the experiments where the participants just applied one (i.e., the between-subjects experiments). Table \ref{summary_subgroup_learning} summarizes the results of the sub-group meta-analysis that we undertook. Figure \ref{forest_learning} shows the corresponding forest plot. Table \ref{marginal_means_learning} shows the estimated quality scores of learning effects divided by development approach. 

\begin{table}[h!]
\small
\begin{center}
\caption{Sub-group meta-analysis: within vs. between.}
\label{summary_subgroup_learning}
\begin{tabular}{llccc} \hline \hline
\textbf{Group} & \textbf{N} & \textbf{Estimate} & \textbf{95\% CI} & \textbf{$I^2$} \\ \hline
Within & 8 & -7.66 & (-18.36, 3.04) & 78.22\% \\
Between & 4 & -4.89 & (-12.5, 2.71) & 0\% \\
Difference & - & -2.76 & (-15.89, 10.36)  & - \\ \hline
\end{tabular}
\end{center}
\end{table}

\begin{table}[!h] \centering 
  \caption{Marginal means for quality: learning effects by treatment.}
  \label{marginal_means_learning} 
\begin{tabular}{llcc} \hline \hline 
\textbf{Treatment} & \textbf{Group} & \textbf{Estimate} & \textbf{95\% CI}\\ \hline 
\multirow{2}{*}{ITL} & Within & 44.38 & (32.08, 56.69) \\
 & Between & 40.78 & (24.34, 57.24)  \\ \hline
 \multirow{2}{*}{TDD}& Within & 36.54 &  (30.51, 42.58) \\
 & Between & 35.98 & (28.20, 43.75) \\ \hline
\end{tabular} 
\end{table}

As we can see in Table \ref{summary_subgroup_learning}, Figure \ref{forest_learning}, and Table \ref{marginal_means_learning}, the sub-group formed by the within-subjects experiments provides less optimistic results for TDD ($M=-7.66$, 95\% \textit{CI}=(-18.36, 3.04)) than the sub-group formed by the between-subjects experiments ($M=-4.86$, 95\% \textit{CI}=(-12.5, 2.71)). However, as the difference of results between both sub-groups is small ($M=-2.76$) and  TDD was applied in the second session in seven out of the eight within-subjects experiments---and, thus, we would expect more optimistic results for TDD---we hypothesize that \textit{learning effects did not materialize}. However, this needs further study. Besides, whereas the sub-group of within-subjects experiments still provides largely heterogeneous results, the sub-group of between-subjects experiments provides homogeneous results. Again, this may be the result of the low number of between-subjects experiments in our family, leading to low $I^2$ statistic accuracy in this sub-group. 
\\

\noindent\fbox{
  \parbox{11.5cm}{
    \textbf{Key findings} \\
    The participants do not appear to have learned anything during the first session that boosted their performance in the second session.
    }
}

\subsection{Order of Treatment Application}
\label{analysis_order_application}

\begin{figure*}[t!]
    \caption{Forest plot: order effects.}
    \label{forest_order}
    \centering
    \includegraphics[width=12cm,keepaspectratio]{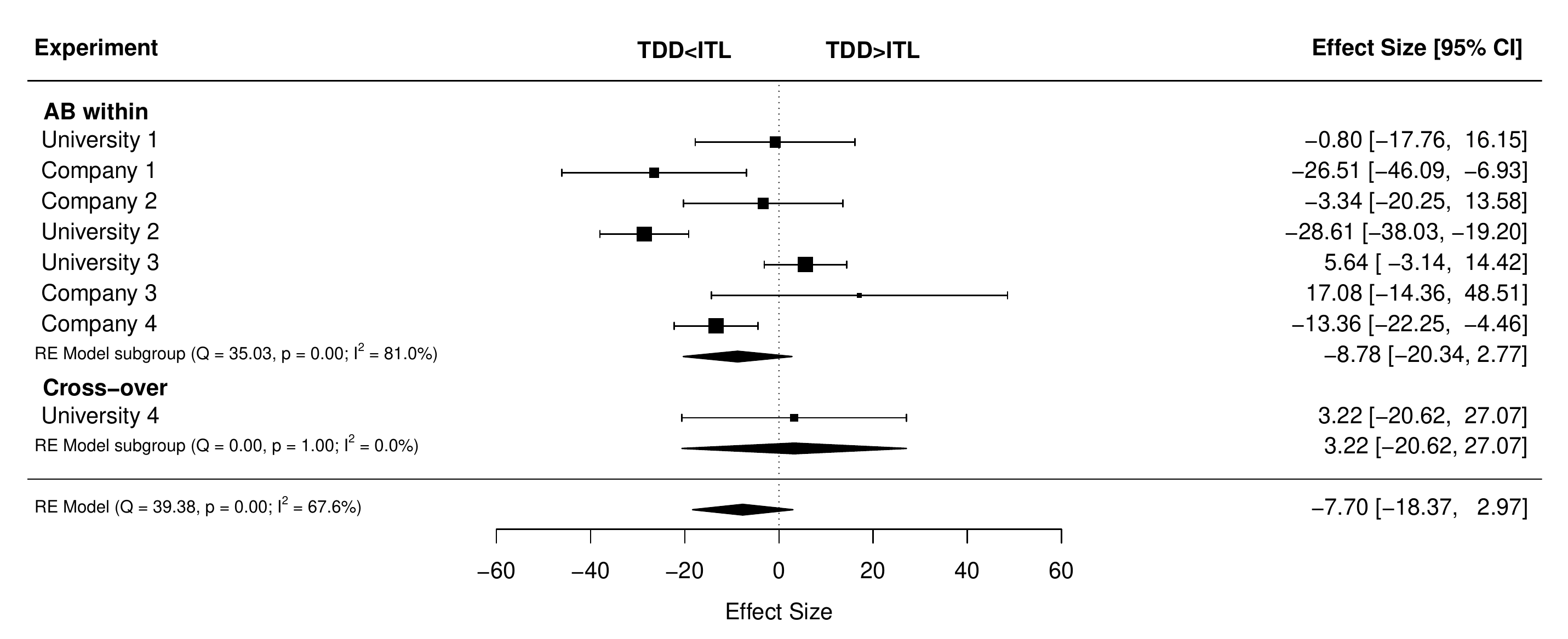}
\end{figure*}

In the discussion of the design of our experiments, we hypothesized that the order of application of ITL and TDD may affect the results. In particular, if the participants applied ITL before TDD, the participants may learn something during the application of ITL (e.g., how to develop in small iterations) that may potentially boost their performance later in the TDD session. In order to check the extent to which this may have influenced the results, we ran a crossover experiment (i.e., an experiment in which participants apply TDD and ITL in a different order) within our family. If the results of this experiment turned out to be similar to the findings of the other within-subjects experiments, then the order of application of TDD and ITL should not be affecting results. Table \ref{summary_subgroup_order} summarizes the results of the sub-group meta-analysis that we undertook to assess the influence of the order of application (i.e., AB within-subjects experiment vs. crossover experiment) on results. Figure \ref{forest_order} shows the corresponding forest plot. Table \ref{marginal_means_order} shows the estimated quality scores of the different orders of application divided by development approach.

\begin{table}[h!]
\small
\begin{center}
\caption{Sub-group meta-analysis: AB Within vs. Crossover.}
\label{summary_subgroup_order}
\begin{tabular}{llccc} \hline \hline
\textbf{Group} & \textbf{N} & \textbf{Estimate} & \textbf{95\% CI} & \textbf{$I^2$} \\ \hline
AB Within & 7 & -8.73 & (-20.33, 2.86) & 81.13\% \\
Cross & 1 & 3.22 & (-20.62, 27.06) & - \\
Difference & - & 11.96 & (-14.55, 38.47)  & - \\ \hline
\end{tabular}
\end{center}
\end{table}

\begin{table}[!h] \centering 
  \caption{Marginal means for quality: order of treatment.} 
  \label{marginal_means_order} 
\begin{tabular}{llcc} \hline \hline 
\textbf{Treatment} & \textbf{Group} & \textbf{Estimate} & \textbf{95\% CI}\\ \hline 
\multirow{2}{*}{ITL} & AB Within & 45.54 & (27.53, 63.54) \\
 & Crossover & 38.20 & (-7.87, 84.28)  \\ \hline
 \multirow{2}{*}{TDD}& AB Within & 36.27 &  (25.58, 46.96) \\
 & Crossover & 41.42 & (14.29, 68.55) \\ \hline
\end{tabular} 
\end{table}

As we can see in Table \ref{summary_subgroup_order}, Figure \ref{forest_order}, and Table \ref{marginal_means_order}, the results of the crossover experiment ($M=3.22$, 95\% \textit{CI}=(-20.62, 27.06)) are way more optimistic than for the AB within-subjects experiments ($M=-8.73$, 95\% \textit{CI}=(-20.33, 2.86)). However, this result should be regarded with caution as we conducted only one crossover experiment (i.e, University 4), and the results of two AB within-subjects experiments (i.e., University 3 and Company 3, see the forest plot in Figure \ref{subgroup_type_of_participant}) were even more optimistic than the findings of the crossover experiment. Finally, there is a large heterogeneity of results in the sub-group of AB within-subjects experiments. This may be due to the participation of professionals and students with different levels of experience across the experiments.
\\

\noindent\fbox{
  \parbox{11.5cm}{
    \textbf{Key findings} \\
        The order of application of the development approaches may be affecting results. However, more crossover experiments are needed to assess this issue.
    }
}

\subsection{Discussion of Moderators at Experiment Level}
\label{findings_characteristics}

We learned that the professionals achieved higher quality than the students with both ITL and TDD in our family. This finding was at odds with the results of the self-assessment questionnaires that we handed out to participants, where most students classed themselves as being at least as experienced as professionals. This observation may be related to the Dunning-Kruger effect \cite{kruger1999unskilled}. Other SE community members (e.g., \cite{baltes2018towards,jung2013experimental}) have observed this effect before. In short, the Dunning-Kruger effect suggests that the knowledge limitations of the respondents---in this case students---may bias their perceptions when rating their skills, which may lead to over-optimistic self-assessments. We hypothesize that this may have led the students to rate themselves as more experienced than professionals---despite their poorer performance with both TDD and ITL. In view of the results in our family of experiments and of others in the SE community \cite{baltes2018towards,jung2013experimental}, self-assessment questionnaires---contrary to what Feigenspan et al. reported \cite{feigenspan2012measuring}---may, therefore, be unsuitable for describing experience within heterogeneous populations. However, we put this forward as a hypothesis and suggest that further studies be conducted to assess the extent to which this hypothesis holds. Additionally, we call for further research aiming to develop new measuring instruments for capturing participant experience in SE empirical studies. 

Although professionals achieved higher quality than students with both ITL and TDD, the performance of professionals with TDD compared to ITL dropped more than for students. This may suggest either that students learn completely new development approaches---such as TDD---faster than professionals (and, thus, students manage to minimize their expected losses when applying a completely new development approach) or that professionals have higher opportunity costs---because they achieve higher quality scores with ITL than students---and, therefore, the drop in their performance when applying TDD is greater than for students. Again, these findings are at odds with outcomes published in the TDD literature, where professionals usually benefit most from TDD. We put this contradictory evidence in the literature down to either the longer duration of the studies in which professionals participate---who, as a result, have more time to get used to TDD---or the level at which quality is measured in these studies (i.e., quality measured at team level at the end of system development rather than for solo programmers). In any case, our recommendation is to run \textit{longitudinal studies with students and professionals to learn whether it is participant type}---or other participant-related characteristics like age---\textit{that are behind the different learning rates (or opportunity costs)} or \textit{experiments with TDD experts to learn whether, after enough exposure, TDD significantly outperforms control approaches}.   

We also learned that the programming environment affected the performance of both ITL and TDD similarly in our experiments---and thus, the difference in performance between ITL and TDD was unaffected by the programming environment. In other words, the different programming environments did not affect the results of the experiments that we ran. In view of this, we recommend that \textit{the programming environments be adapted to the requirements of the host institutions when conducting experiments}. This should increase the external validity of the results (as TDD will be evaluated in potentially more diverse programming environments), and the internal validity of the results (as participant unfamiliarity with the programming environment used could interfere with their ability to apply the development approaches). Three out of four experiments using the C programming environment led to negative results in our family. This observation is at odds with the results typically achieved in case studies, where C environments are only associated with positive results. These conflicting results may be due to the many variables on which such case studies and our experiments differ.

We ended by assessing the extent to which the learning effects from one session to another or the order of application of the development approaches affected the results of our experiments. We learned that the results of the experiments were unaffected by whether the participants applied both ITL and TDD or just one of the development approaches. In other words, no learning effects appeared to materialize in our experiments. In view of this, we recommend that \textit{within-subjects instead of between-subjects experiments be run to evaluate the performance of TDD}. This increases both the external validity of results (as more than one task needs to be exercised in within-subjects experiments) and the precision of results (as the participants are assessed against themselves, which reduces the variability of the results \cite{cumming2013understanding,field2013discovering}). It also accommodates the above recommendation related to the effects of the task being developed on results. Finally, we noticed that the order of application of TDD (i.e., in the first session, or in the second session) \textit{may be affecting results}. However, as we only ran one crossover experiment within our family, and as two of the AB within-subjects experiments provided even more extreme results than the crossover experiment, we recommend that \textit{more crossover experiments be run to assess the extent to which the order of TDD application affects the results}.

Finally, we found that there was still a large heterogeneity of results  in almost all of the sub-groups that we analyzed (except the sub-group composed exclusively of between-subjects experiments with students). This suggests either that there are more experiment-level characteristics impacting the results \textit{within each sub-group} or that \textit{the different participant characteristics across the experiments} are responsible for this heterogeneity of results. Unfortunately, further sub-dividing the sub-groups according to other experiment-level characteristics would affect the reliability of the resulting estimates (as fewer experiments would contribute towards the joint result in each sub-group \cite{borenstein2011introduction}). Thus, we think that the only option open to try to explain the heterogeneity of results would be to study how participant characteristics impacted TDD performance. However, as we changed the scales of the questionnaires in our experiments---and our questionnaires may not have accurately captured the different participant experience either---we are unable to study this in our family. We suggest that \textit{a family of experiments be run with improved instruments for measuring this characteristic in the future}.

\section{RQ4: Participant-Level Moderators}
\label{rq4}

Throughout this section, we answer RQ4: To what extent is quality affected by the participant characteristics: programming, programming language, unit testing or testing tool experience? Note that another reason for conducting families of experiments is to hypothesize on participant-level moderators that may be influencing the results \cite{adriguidelines}. For this purpose, we assess the influence of participant characteristics for professionals and students separately in Sections \ref{analysis_professionals} and \ref{analysis_students}, respectively. Then, in Section \ref{findings_participants}, we discuss and contextualize our findings.

\subsection{Professional Characteristics}
\label{analysis_professionals}

\begin{figure}[h!]
    \caption{Profile plot: mean difference in quality (TDD-ITL) for professionals per experience level in Questionnaire 1.}
    \label{profile_plot_q1_professionals}
    \centering
    \includegraphics[width=9cm,keepaspectratio]{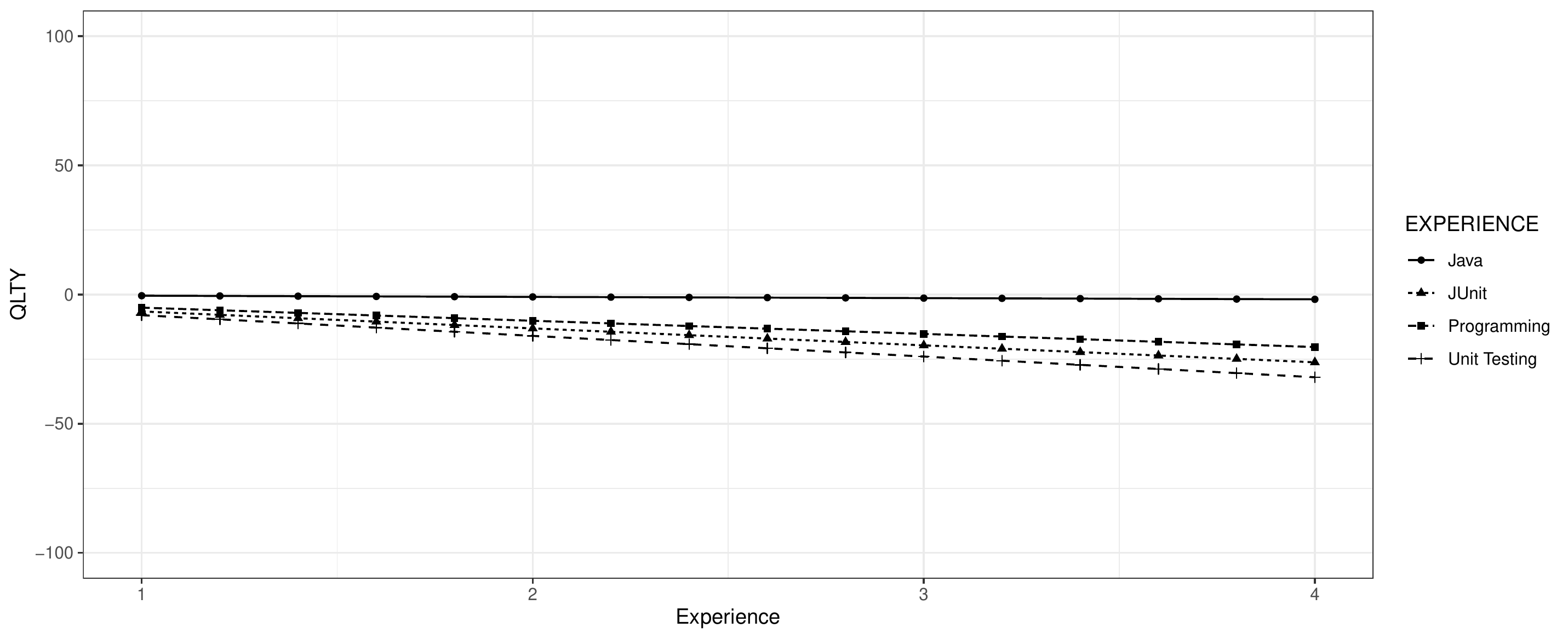}
\end{figure}

\begin{figure}[h!]
    \caption{Profile plot: mean difference in quality (TDD-ITL) for professionals per experience level in Questionnaire 2.}
    \label{profile_plot_q2_professionals}
    \centering
    \includegraphics[width=9cm,keepaspectratio]{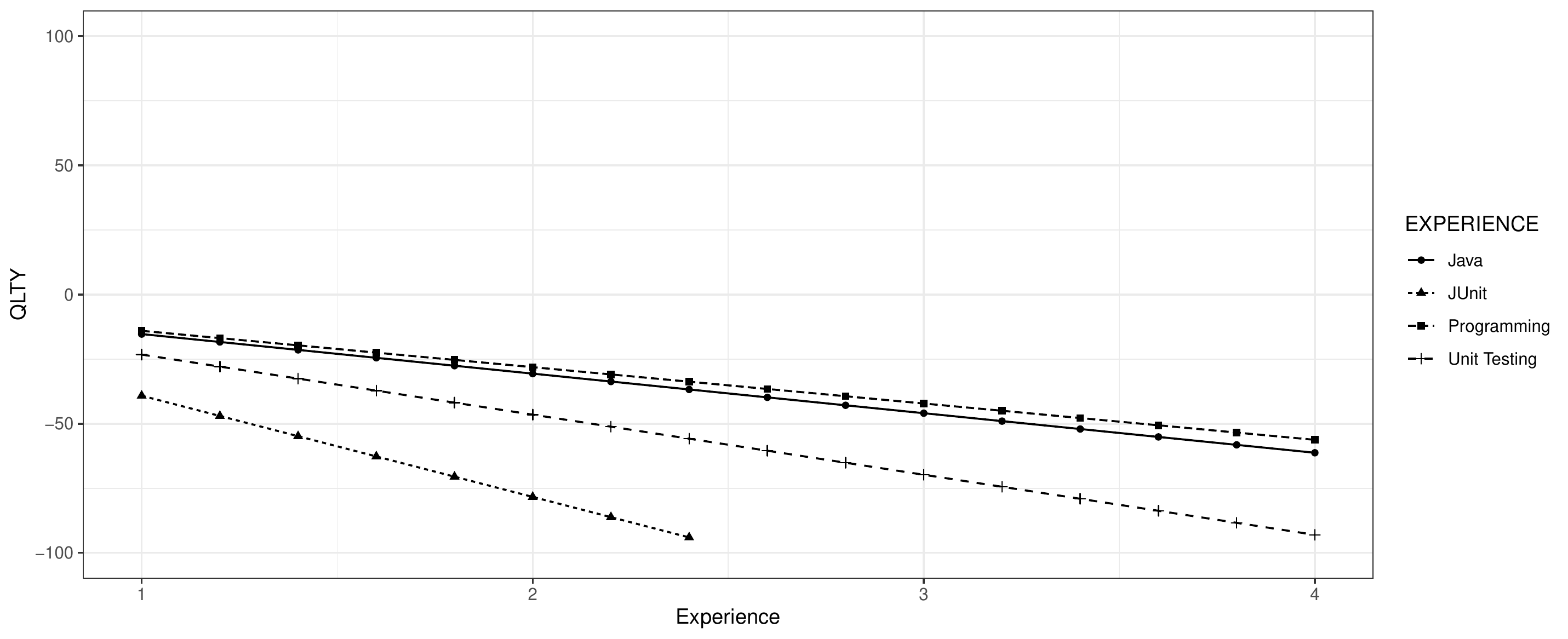}
\end{figure}

Tables \ref{summary_questionnaire_one_professionals} and \ref{summary_questionnaire_two_professionals} show the ANOVA p-values and marginal means for the interaction between treatment and each participant characteristic in Questionnaire 1 (37 participants) and Questionnaire 2 (17 participants). Figures \ref{profile_plot_q1_professionals} and \ref{profile_plot_q2_professionals} show the corresponding profile plots. Note that downward-sloping lines in the profile plots indicate greater experience and lower quality for TDD with respect to ITL (the opposite applies for upward-sloping lines).

\begin{table}[h!]
\caption{Treatment by participant experience interaction: Questionnaire 1 for professionals.}
\label{summary_questionnaire_one_professionals}
\begin{tabular}{llll}
\hline \hline 
\textbf{Characteristic} & \textbf{Estimate} & \textbf{95\% CI} & \textbf{p-value} \\ \hline
Treatment:Programming & -5.08 & (-21.55, 11.38) & 0.53 \\
Treatment:Prog. Language & -0.47 & (-16.82, 15.89)  & 0.95 \\
Treatment:Unit Testing   & -8.01 & (-29.22, 13.20)  & 0.45  \\
Treatment:Testing Tool   & -6.56 & (-55.70, 42.58)  & 0.79 \\  \hline         
\end{tabular}
\end{table}

\begin{table}[h!]
\caption{Treatment by participant experience interaction: Questionnaire 2 for professionals.}
\label{summary_questionnaire_two_professionals}
\begin{tabular}{llll} \hline \hline
\textbf{Characteristic} & \textbf{Estimate} & \textbf{95\% CI} & \textbf{p-value} \\ \hline
Treatment:Programming    & -14.07 & (-39.73, 11.59)  & 0.26 \\
Treatment:Prog. Language & -15.32 & (-30.27, -0.37)  & 0.05 \\
Treatment:Unit Testing   & -23.26 & (-44.81, -1.72)  & 0.04 \\
Treatment:Testing Tool   & -39.18 & (-61.83, -16.55) & 0.00 \\ \hline         
\end{tabular}
\end{table}

The exploratory analyses of the data for professionals suggest that professionals with more programming, programming languages, unit testing, and testing tools experience perform worse with TDD compared to ITL. In order words, the more experience professionals have, the larger the drop in quality with TDD compared to ITL is. This holds in both questionnaires. \\

\noindent\fbox{
  \parbox{11.5cm}{
    \textbf{Key findings} \\
    The more programming, programming languages, unit testing, and testing tool experience professionals have, the worse they perform with TDD in comparison to ITL.
    }
}

\subsection{Student Characteristics}
\label{analysis_students}

\begin{figure}[h!]
    \caption{Profile plot: mean difference in quality (TDD-ITL) for students per experience level in Questionnaire 1.}
    \label{profile_plot_q1_students}
    \centering
    \includegraphics[width=9cm,keepaspectratio]{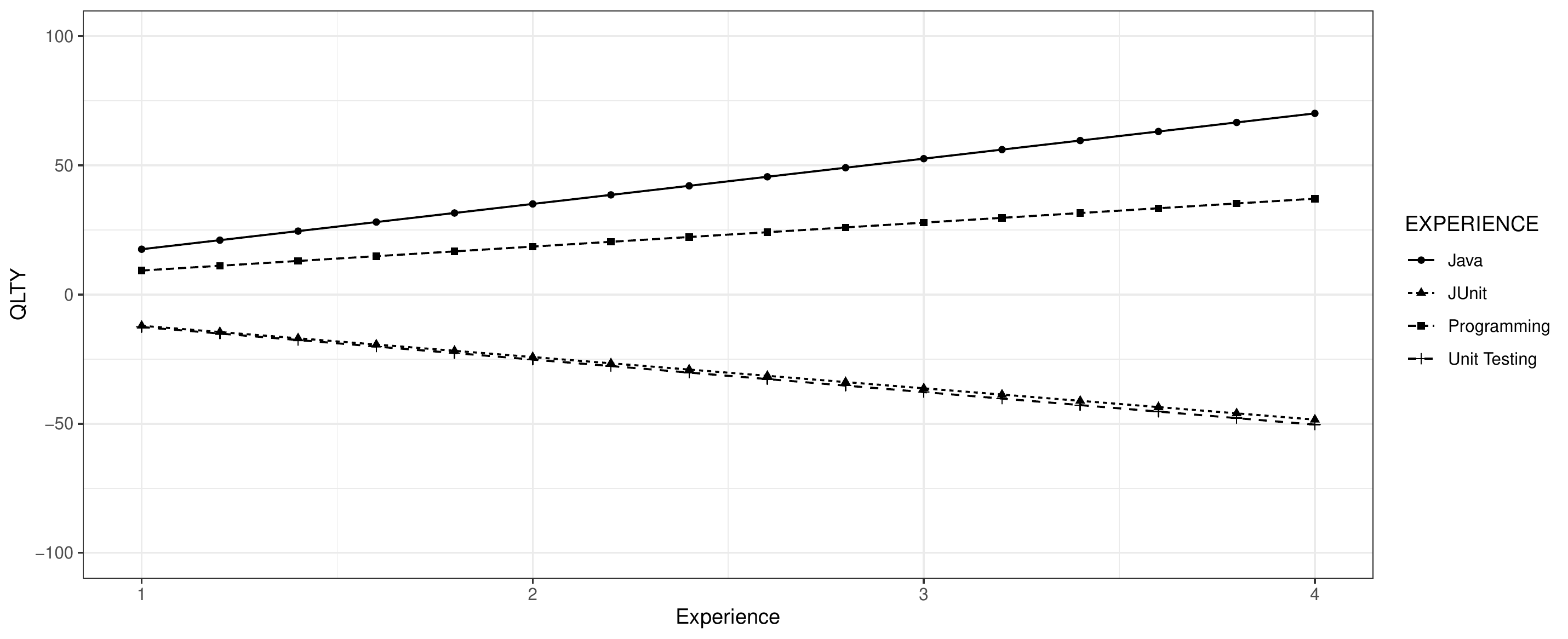}
\end{figure}

\begin{figure}[h!]
    \caption{Profile plot: mean difference in quality (TDD-ITL) for students per experience level in Questionnaire 2.}
    \label{profile_plot_q2_students}
    \centering
    \includegraphics[width=9cm,keepaspectratio]{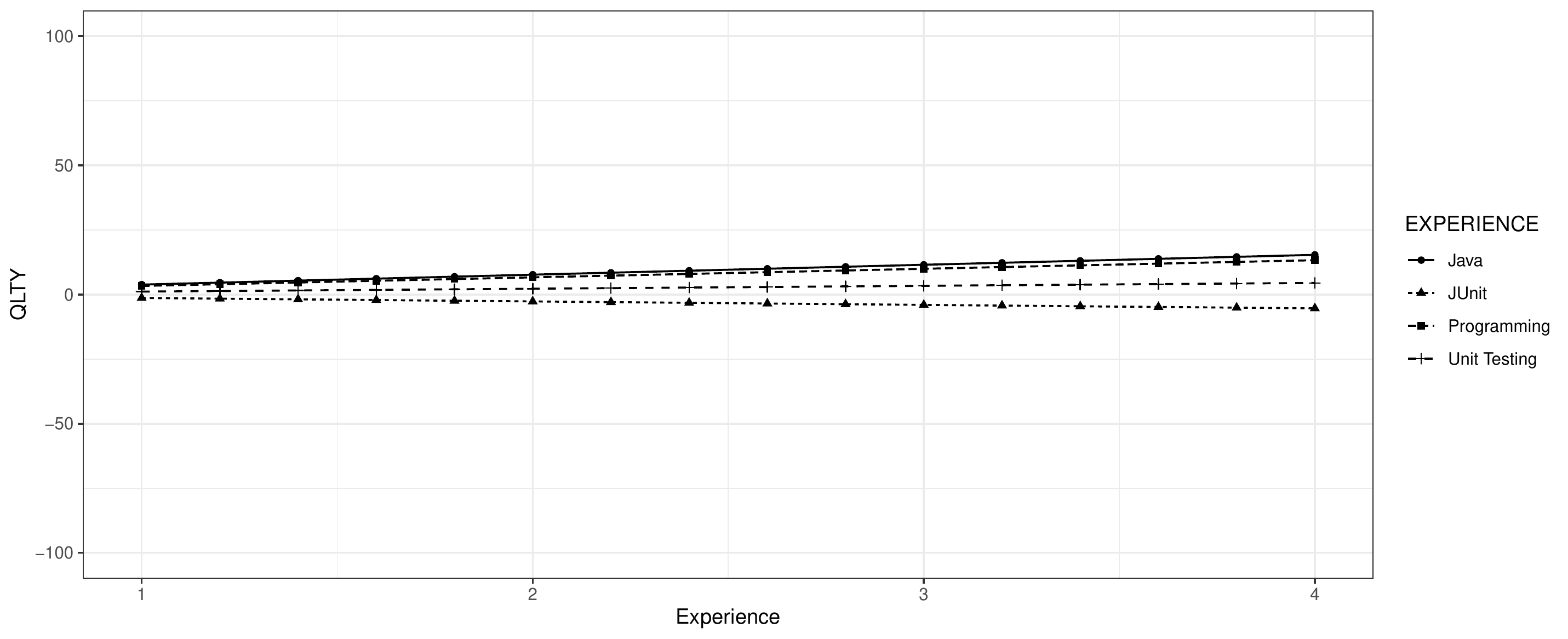}
\end{figure}

Tables \ref{summary_questionnaire_one_students} and \ref{summary_questionnaire_two_students} show the ANOVA p-values and marginal means for the interaction between treatment and each participant characteristic in Questionnaire 1 (73 participants) and Questionnaire 2 (261 participants). Figures \ref{profile_plot_q1_students} and \ref{profile_plot_q2_students} show the corresponding profile plots.

\begin{table}[h!]
\caption{Treatment by participant experience interaction: Questionnaire 1 for students.}
\label{summary_questionnaire_one_students}
\begin{tabular}{llll} \hline \hline
\textbf{Characteristic} & \textbf{Estimate} & \textbf{95\% CI} & \textbf{p-value} \\ \hline
Treatment:Programming    & 9.27   & (-12.49, 31.04)  & 0.39 \\
Treatment:Prog. Language & 17.53  & (-3.22, 38.28)   & 0.10 \\
Treatment:Unit Testing   & -12.59 & (-44.52, 19.33)  & 0.43 \\
Treatment:Testing Tool   & -12.11 & (-46.61, 22.39)  & 0.48 \\ \hline           
\end{tabular}
\end{table}

\begin{table}[h!]
\caption{Treatment by participant experience interaction: Questionnaire 2 for students.}
\label{summary_questionnaire_two_students}
\begin{tabular}{llll} \hline \hline
\textbf{Characteristic} & \textbf{Estimate} & \textbf{95\% CI} & \textbf{p-value} \\\hline
Treatment:Programming    & 3.32  & (-5.83, 12.47)   & 0.47 \\
Treatment:Prog. Language & 3.84  & (-5.77, 13.45)   & 0.43 \\
Treatment:Unit Testing   & 1.12  & (-10.01, 12.25)  & 0.84 \\
Treatment:Testing Tool   & -1.34 & (-13.33, 10.65)  & 0.82  \\ \hline          
\end{tabular}
\end{table}

The exploratory analyses of students tell a different story. In particular, greater programming-related (programming and programming language) experience tends to benefit TDD. This holds for both questionnaires.\\

\noindent\fbox{
  \parbox{11.5cm}{
    \textbf{Key findings}
    \begin{itemize}
        \item{More programming-related experience tends to benefit TDD.}
        \item{More testing-related experience tends to benefit ITL. }
    \end{itemize} 
    }
}

\subsection{Discussion of Moderators at Participant Level}
\label{findings_participants}

As for professionals, greater testing-related (unit testing and testing tool) experience tends to benefit ITL. The results are similar for both questionnaires---although more unit testing experience in Questionnaire 2 has a negligible positive effect on TDD. We hypothesize that this may be because professionals may be seasoned users of test-last approaches---more in line with waterfall model development. Thus, they may find it easier to continue with their usual development habits and more difficult to break with them and move towards new---and perhaps less intuitive---test-first development approaches (like TDD). It is as if their previous programming habits lead to a higher resistance to change---at least when applying the new approach in a single experimental session.

On the contrary, when the students apply TDD rather than ITL, quality drops are comparatively smaller than for professionals. We hypothesize that this may be because students with previous programming experience may be more flexible than professionals about learning completely new development approaches (such as TDD). In fact, they may have not acquired such strong development habits as  professionals and, in this sense,  may be less resistant to change. In this regard, students with a lot of programming experience may be benefiting from what \cite{beck2003test} referred to as TDD's \textit{virtuous cycles}: creating a test first and then implementing a small functionality to pass that test may increase developer focus and, possibly, performance. However, this is more noticeable for students who have quite a lot more programming experience than their peers.
 
We hypothesize that the negative influence of student testing experience on TDD performance may be because students with more testing experience may be more used to creating tests after seeing the implemented code. The reverse development approach applied in TDD may, especially if students are short on programming experience, impinge on their ability to implement new tests. After all, designing a test for an as yet unimplemented functionality is likely to slow down their progress if they do not have a clear idea of how it should be implemented. This may help to explain why students with a lot of testing experience benefit more from use of ITL---as, with ITL, they implement the code first, which facilitates test development (and matches their programming approach).

Finally, students' lower resistance to change---and, possibly, in view of our observations, greater motivation or flexibility to learn new development approaches---may have led them to minimize their quality losses when applying TDD over ITL with respect to professionals.

As these analyses are exploratory, we put forward these claims as mere hypotheses. We note that the effect of participant experience on the performance difference between TDD and ITL tends to be larger with smaller sample sizes for both students and professionals (i.e., with Questionnaire 1---attaching years of experience to experience levels---for students, and with Questionnaire 2---not attaching years of experience to experience levels---for professionals). This may be because there are influential points (i.e., points with abnormally high X values---in our case, reported experience levels) that pull the regression lines towards each other, especially if sample size is small \cite{quinn2002experimental}. In the case of students, this may happen because regression lines are pooled towards the results of only a few participants declaring $>$10 years of experience. In the case of professionals, this may happen because the regression lines are pooled towards the results of only a few participants declaring large subjective experiences---as we observed that most professionals tended to be conservative when reporting their experience.

Finally, we find that the influence of experience on results flattens out with larger sample sizes (as with Questionnaire 2 and students, where 261 participants were analyzed). This may suggest that either: (1) experience levels influence the performance achieved with ITL or TDD similarly when results are more accurate \cite{cumming2013understanding}; or (2) the self-assessment questionnaires that we used did not capture participant experience---as when sample sizes get larger, the influence of the experience on results flattens out as random noise.

\section{Threats to validity}
\label{threats_to_validity}

\subsection{Statistical Conclusion Validity}
\label{conclusion_threats}

We relied upon parametric statistical tests (i.e., $t$-tests and linear mixed models \cite{brown2014applied,cumming2013understanding,field2013discovering}) to analyze the data of our family of experiments. We relied upon the robustness of the $t$-test to departures from normality when analyzing the data of our experiments \cite{schmider2010really,vickers2005parametric,de2013using}. To double-check, we have also analyzed the data of the experiments with non-parametric tests (i.e., the Wilcoxon signed rank test for within-subjects experiments, and the Wilcoxon rank sum test for between-subjects experiments \cite{field2013discovering}), and computed the corresponding non-parametric effect sizes for each individual experiment (i.e., the point biserial correlation following \cite{rosenthal1991effectsizes}), and their variances. Finally, we pooled the effect sizes---and their respective variances---using meta-analysis. The observed results are consistent with parametric test outcomes. 

We relied upon the central limit theorem to analyze the data of all the experiments together with linear mixed models \cite{lumley2002importance}. We ensured the robustness of the results that we provided by meta-analyzing the data with both one-stage and two-stage IPD models \cite{riley2010meta}. In the spirit of open-science and to ensure the transparency of results, we uploaded the experimental data of the family and the statistical analyses that we ran to a GitHub repository. 

We acknowledge that the exploratory analyses made on participant experience may increase the risk of committing inflated type I error rates. To avoid this problem, we focus more on the magnitude and sign of moderator effects---and their corresponding 95\% CIs---than on their statistical significance.

\subsection{Internal Validity}
\label{internal_threats}

We acknowledge that, as our experiments were designed merely to prove that the development approach caused an effect on quality (as we just randomized participants to tasks and treatments), we are not able to draw any cause-effect relationships for the other variables that we investigated \cite{lau1998summing} (e.g., participant type, programming environment, etc.). This is because differences across experiment results could have been caused either by the variable under investigation (e.g., the participant type) or unacknowledged variables confounded with the investigated variable (e.g., professionals were older than students, and, thus, participant age could be the real cause behind the differences in the results). Still, we think that these analyses may serve to foster further research and as a proxy for understanding the heterogeneity of results reported in the TDD literature. 

We acknowledge that, as our experiments are embedded within training courses (which meant that no participant had any previous experience), there is a selection bias.

We acknowledge that the distinction between students and professionals is controversial, and this could have an impact on our results (as some students may also be working in industry, and some professionals may also be studying) \cite{falessi2018empirical,sjoberg2018students}.

We acknowledge that different test oracles may lead to different results, and therefore the chosen test oracle may be impacting results.

We also acknowledge that, although the IDEs and testing tools used with C++ and C\# are different, we make the simplification of considering them as being a part of the same group of technologies in this study.

We acknowledge that the Hawthorne effect\footnote{A type of reactivity in which individuals modify an aspect of their behavior in response to their awareness of being observed.} might have occurred. However, it would affect TDD and ITL equally.

\subsection{Construct Validity}
\label{construct_threats}

We acknowledge that evaluating the source code with tests not written alongside the code might influence results. However, this applies to both ITL and TDD.

We acknowledge that, although we followed similar approaches to those suggested in SE \cite{falessi2018empirical,feigenspan2012measuring} for designing our experience questionnaires, the questionnaires may not be accurately capturing the participant characteristics. We also acknowledge that the change of scales in some experiments did not help during the meta-analysis phase. However, we provide testimony that self-assessment questionnaires may be unsuitable for assessing the characteristics of heterogeneous populations and also call for further research aiming to develop new measurement instruments in order to evaluate participant characteristics in joint analyses in the future. A possible solution could be to ask participants about the number of programming courses they have taken, the number of projects they have worked on, or how many years they have been working in industry.

 Notice that we do not check treatment conformance. Participants may have applied a hybrid approach instead of ITL and TDD, which may have affected the construct validity of our studies. However, none of the empirical studies on TDD check treatment conformance either.

Additionally, other issues, such as the average development cycle duration, duration uniformity and refactoring effort, rather than the code/test writing order could pose a challenge \cite{karac2018we}.

Finally, we acknowledge that the \textit{hypothesis guessing} threat might have materialized \cite{wohlin2012experimentation}. People taking part in an experiment might try to figure out what the purpose and intended result of the experiment is. They are then likely to base their behavior on their guesses, acting either positively or negatively depending on their attitude towards the anticipated hypothesis.

\subsection{External Validity}
\label{external_threats}

Due to space restrictions, we focused on \textit{external quality} only throughout this article, omitting other quality attributes (e.g., internal quality, maintainability, etc. \cite{iso25010}). We selected external quality as it is one of the attributes that---according to its proponents \cite{astels2003test,beck2003test}---TDD benefits most. Also it is one of the most studied attributes so far in the TDD literature \cite{bissi2016effects,causevic2011factors,kollanus2010test,makinen2014effects,munir2014considering,rafique2013effects,shull2010we}, thus allowing for further comparisons of results. Having said this, we acknowledge that our findings are limited exclusively to external quality. Therefore, our results may not be representative of what may happen with other quality attributes. 

As usual in SE experiments \cite{wohlin2012experimentation}, we had to rely on toy tasks to evaluate the performance of TDD and ITL. We relied on toy tasks similar to the ones used in \citeP{erdogmus2005effectiveness},\cite{dieste2017empirical,fucci2017dissection,tosun2017industry}. The downside of the use of toy tasks is a loss of realism, where our results may not be representative of what may happen when developing real industrial projects. Although this is a potential threat to external validity (effects might not be apparent on such small sized programs), it has the advantages of increasing internal validity like:

\begin{itemize}
	\item The possibility of observing task development within a controlled environment and thereby ensuring that external factors do not affect the achieved quality.
	\item The possibility of comparing the results achieved with both TDD and ITL on an identical experimental task (since both TDD and ITL are applied to code an identical experimental task, say, BSK). In turn, this helps to ensure that differences across treatment results are not due to the different experimental tasks being developed with each treatment (as is the case if both treatment and task go hand in hand---e.g. Waterfall Model-System 1 vs. TDD-System 2, as is typical in industrial case studies).
	\item The possibility of comparing the results from industrial experiments and academic experiments---as if professionals had developed industrial tasks and students, toy tasks, the difference in the results across experiments could have been caused either by participant type (i.e., professionals vs. students) or by the different task types developed (either industrial tasks or toy tasks).
\end{itemize}

Having said this, we acknowledge that our findings are limited exclusively to toy tasks. Therefore, our findings may not be representative of what may happen when developing real-life industrial systems. However, we expect our results to be representative for professionals and students who are starting to learn the TDD process when developing toy tasks.

We discarded the data for three tasks (i.e., MP, SDK and SS) when studying the effect of task on results. We did this because each task was used in at most two experiments, and we did not have enough data to provide accurate results. We acknowledge that this decision may have limited the external validity of our findings with regard to the effect of the task on results. However, considering that MR and BSK influence both ITL and TDD similarly, we see no reason why MP, SDK or SS should behave differently. Still, this deserves further study.

We acknowledge that, as our experiments are embedded within training courses and participants are novices and not experienced people, the generalization of results is limited to TDD novices only.

\section{Conclusions}
\label{conclusions}

Most of the studies published so far on TDD claim that TDD outperforms control approaches with regard to external quality \cite{bissi2016effects,causevic2011factors,kollanus2010test,makinen2014effects,munir2014considering,rafique2013effects,shull2010we}. However, the extent to which TDD outperforms control approaches seems largely dependent upon a myriad of yet to be discovered variables \cite{offutt2018don}. The disaggregation of the results of the TDD studies published so far according to their research methods (i.e., case studies, surveys, experiments), participant type (i.e., professionals vs. students), task type (toy task vs. industrial project), unit of analysis (i.e., solo programmers, pair programmers, or teams), project length (i.e., days, weeks, months or years), and programming environment (i.e., Java vs. C/C++/C\# technologies) could provide further insight. In particular, we found that, while the most optimistic results for TDD tend to be achieved in case studies conducted over months or years with development teams coding industry-relevant systems, the results achieved in the studies where TDD is evaluated either on solo or pair programmers for toy tasks tend to be more conservative. We put this down to the shorter time frame in which experiments are run compared to case studies. In particular, TDD may not ''click'', and improve external quality, until after a fairly long exposure time. The many changes made across the studies complicate the aggregation of results, and the identification of the variables that are impacting the results.

We adopted a different approach to provide joint results and identify variables impacting results: we conducted a family of experiments composed of four industrial experiments and eight academic experiments. We used identical tasks and response variable operationalizations across all the experiments. We adapted the experimental configurations to the restrictions of our industrial partners. We evaluated the extent to which the shortcomings of these experimental configurations impacted the results by altering the configurations of the academic experiments that we ran and comparing their results. We provided joint results and assessed the potential effects of the experimental changes using meta-analysis \cite{borenstein2011introduction}.

With our family, we learned that TDD novices seem to perform \textit{slightly} better using iterative test-last development (i.e., ITL, the reverse approach of TDD according to Erdogmus et al. \citeP{erdogmus2005effectiveness}) than with TDD in terms of quality. We also learned that the task under development consistently affects---either increasing or decreasing---the quality scores achieved with both TDD and ITL for TDD novices. Finally, we reached the conclusion that there is no 'wonder' task that boosts quality with a particular development approach: whatever affects TDD, also affects ITL, and vice versa.

Then we learned that although professionals achieve higher quality than students with ITL and TDD, their performance with TDD dropped more than for students. We hypothesize that this may be because either professionals and students have different learning rates or professionals have higher opportunity costs when applying TDD. However, further experiments are needed before we can make definite claims regarding this issue---as the difference in performance between TDD and ITL for professionals and students was not statistically significant. Additionally, we learned that self-assessment questionnaires may not be informative for describing the characteristics of heterogeneous populations: the students in our family seemed to be overrating their expertise because they were not sufficiently knowledgeable. Neither the programming environment, nor the learning effects from one development approach to another influenced the results of our experiments. The order of application of TDD (i.e., before or after ITL) may be affecting the results. However, more experiments are needed to assess the extent to which this holds.

In view of our findings, we acknowledge that the differences in results observed between case studies and surveys, on the one hand, and experiments, on the other (i.e., positive vs. negative, respectively), may be influenced by the fact that experiment participants were not previously familiar with the TDD process. Consequently, we call for experiments with TDD experts to be run in order to eventually be able to ascertain whether the optimistic results achieved in case studies with TDD are also reproduced in experimental settings. Based on our findings, we also recommend adapting the programming environments of the experiments to host institution demands, using as few tasks as necessary to evaluate the performance of TDD and the control approach, favoring the execution of within-subjects over between-subjects experiments to assess the performance of TDD, designing new measuring instruments to measure participant experience, and running further studies to study the effect of participant characteristics on results. 

Throughout this article, we argue that, while most empirical studies show that TDD outperforms control approaches (e.g., ITL or waterfall model), the extent of this outperformance appears to be dependent upon a great many variables. In addition, despite the many years of TDD research, little is yet known about how TDD performs in industrial experiments \cite{offutt2018don}. In view of this and the results of our family of experiments, we conclude:

\begin{itemize}
        \item{As the results of the experiments in which the participants applied both ITL and TDD and others in which they applied just one approach were similar, we recommend running within-subjects experiments instead of between-subjects experiments to increase the precision of results.}
        \item{As the programming environment does not appear to influence the results of the experiments, we hypothesize that differences across study results in the literature are unlikely to be caused by the different programming environments.}
        \item{The task under development seems to have a similar influence---either increasing or decreasing---on quality with both ITL and TDD. Thus, as long as the same task is developed with TDD and the control approach in a study, differences across study results are unlikely to be caused by the different tasks being developed. However, as quality seems largely dependent upon the task being developed, we recommend either: (1) the use of only one task in between-subjects experiments so as not to increase the variability of the data and, thus, lower the precision of results; or (2) the execution of within-subjects experiments so as to cancel out the extra variability potentially introduced by different tasks.}
        \item{As a range of variables---different study durations, experiment participant unfamiliarity with TDD, the control approach (e.g., waterfall model) used for the performance comparison with TDD in case studies, or system complexity---may be behind the conflicting results of case studies and surveys, on the one hand, and experiments, on the other (i.e., positive vs. negative, respectively), more experiments investigating these variables are needed to assess these hypotheses.} 
    \end{itemize}

Finally, we have made the experimental data generated by the participants in our family (i.e., their Java and Visual Studio solutions), their raw data (i.e., their external quality scores, treatment to task assignments and experience-level variables) and the R code that led to the results reported throughout this study available in a GitHub repository. Hopefully, this could be used by other SE researchers to run their own studies and round out our conclusions.

\begin{acknowledgements}
This research was developed with the support of project PGC2018-097265-B-I00, funded by: FEDER/Spanish Ministry of Science and Innovation—Research State Agency. We would like to thank the participants in the ESEIL experiments: this research would not have been possible without your help. We would also like to thank the anonymous reviewers for their valuable comments during the review of the manuscript. 
\end{acknowledgements}

\bibliographystyle{spmpsci}      
\bibliography{biblio}   

\section*{Appendix A}
\bibliographystyleP{spmpsci}
\bibliographyP{primary}

%
%

\clearpage

\hfill \break
\begin{floatingfigure}{1.2in}
\includegraphics[width=1in,clip,keepaspectratio]{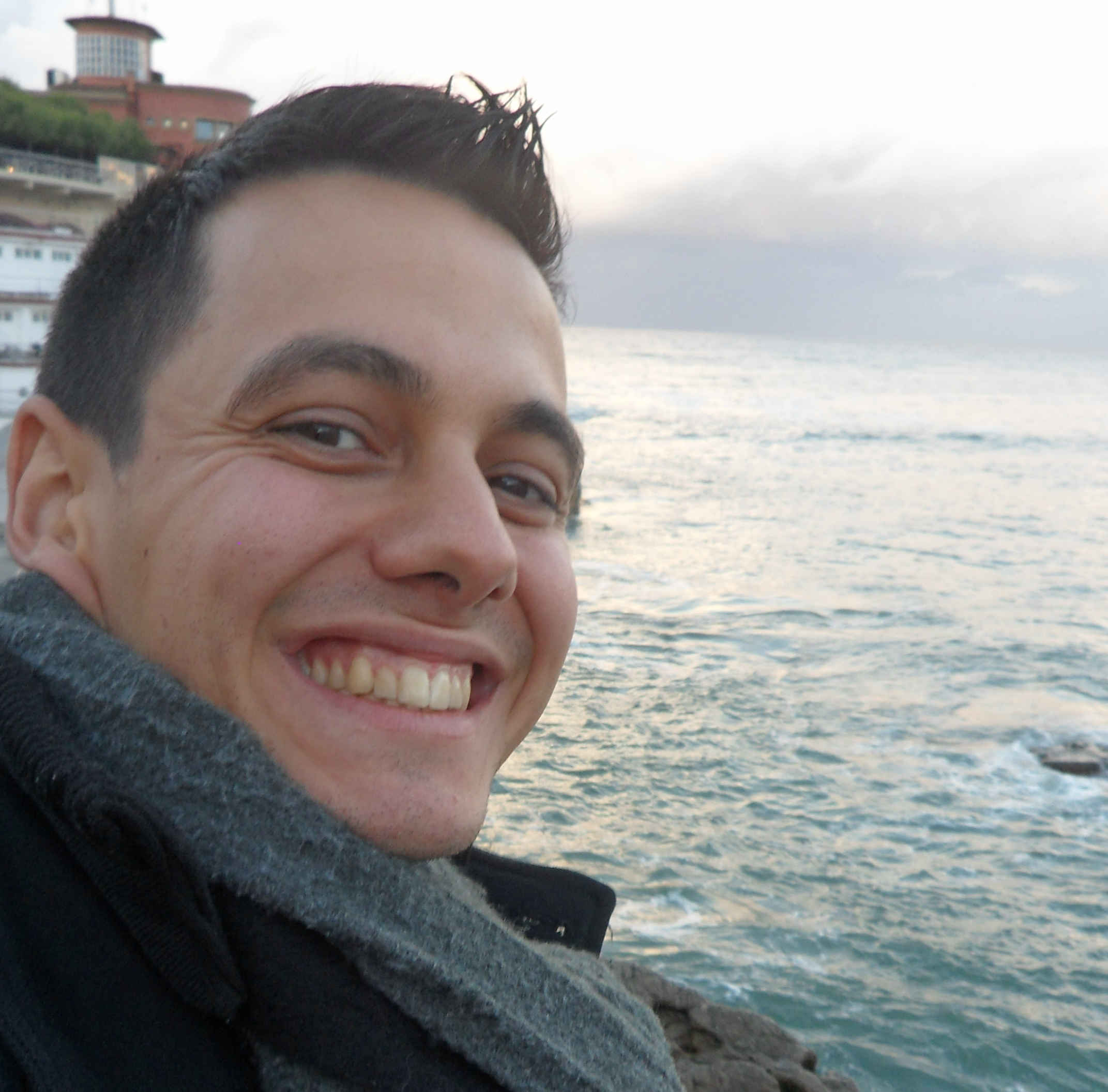}
\end{floatingfigure}
\noindent \textbf{Adrian Santos} received his MSc in Software and Systems and MSc in Software Project Management at the Technical University of Madrid, Spain, and his MSc in IT Auditing, Security and Government at the Autonomous University of Madrid, Spain. He obtained his PhD in Software Engineering at the University of Oulu, Finland. He is currently working as a software engineer in industry. 
\newline

 
\hfill 
\begin{floatingfigure}{1.2in}
\includegraphics[width=1in,clip,keepaspectratio]{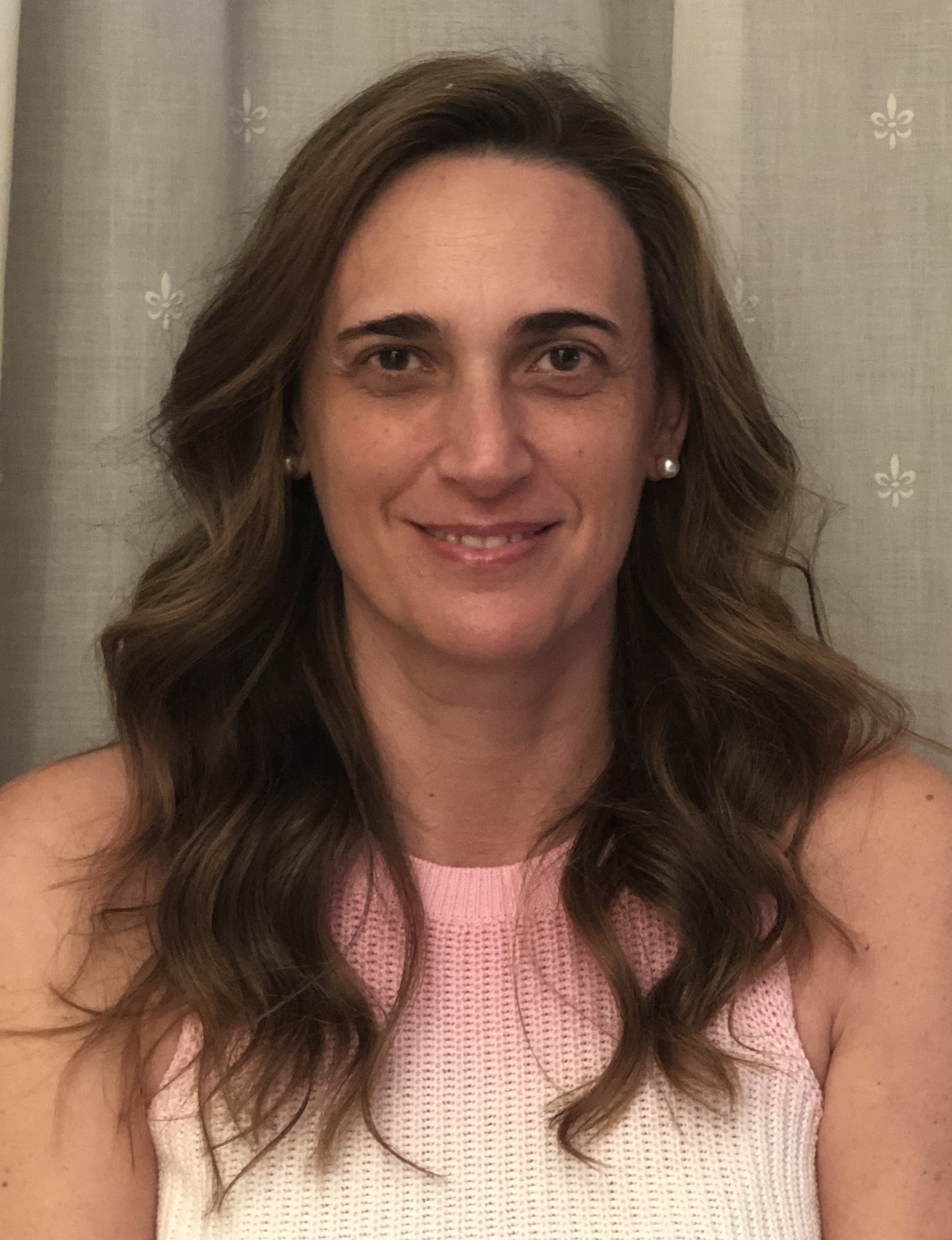}
\end{floatingfigure}
\noindent \textbf{Sira Vegas} has been associate professor of software engineering with the  School of Computer Engineering at the Technical University of Madrid, Spain, since 2008. Sira belongs to the review board of IEEE Transactions on Software Engineering, and is a regular reviewer of the Empirical Software Engineering Journal. She was program chair for the International Symposium on Empirical Software Engineering and Measurement in 2007. 
\newline
 
\vspace{0.2cm}

\hfill 
 \begin{floatingfigure}{1.5in}
 \includegraphics[width=1.3in,clip,keepaspectratio]{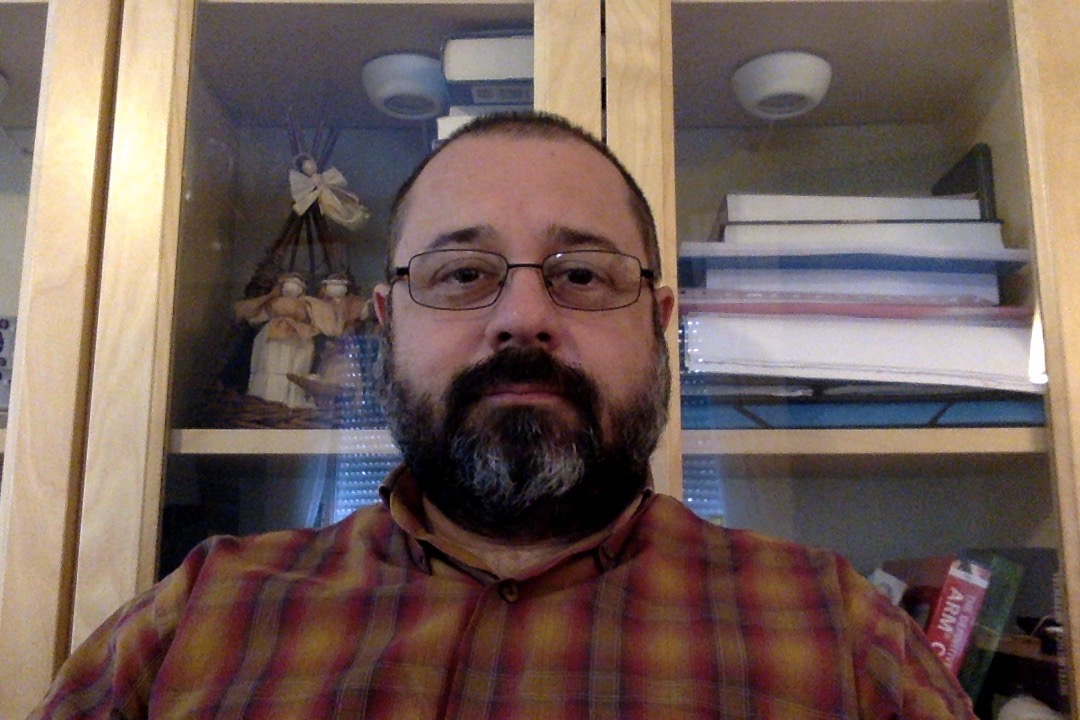}
 \end{floatingfigure}
 \noindent \textbf{Óscar Dieste} received a degree in Computing from the University of La Coruña and his PhD from the University of Castilla-La Mancha. He is a researcher with the UPM’s School of Computer Engineering. He was previously with the University of Colorado at Colorado Springs (as a Fulbright scholar), the Complutense University of Madrid, and the Alfonso X el Sabio University. His research interests include empirical software engineering and requirements engineering.  
\newline


\hfill 
 \begin{floatingfigure}{1.2in}
 \includegraphics[width=1in,clip,keepaspectratio]{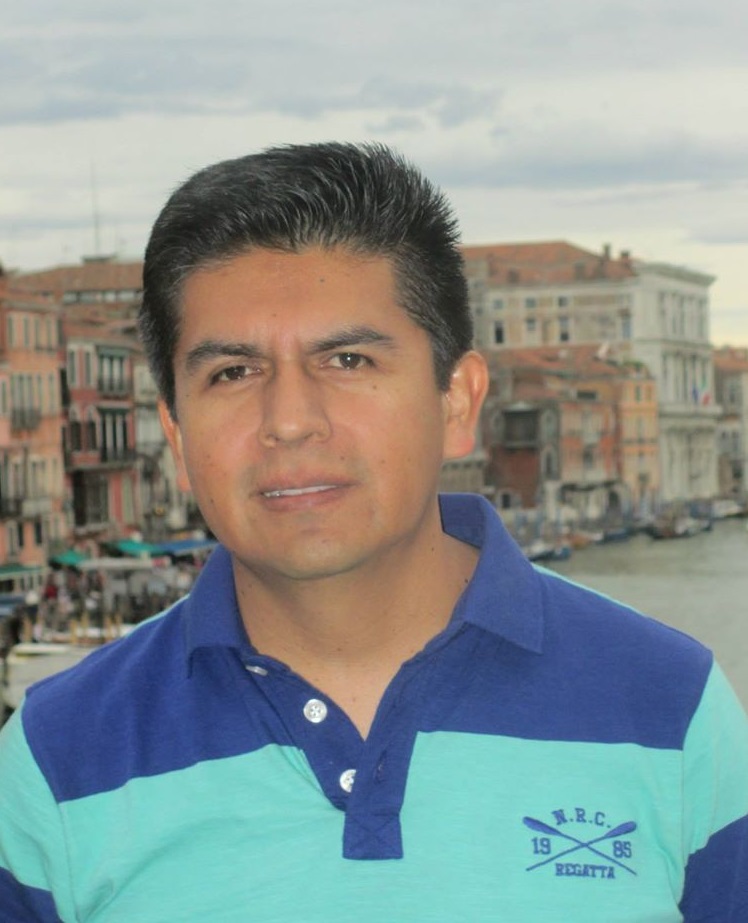}
 \end{floatingfigure}
 \noindent \textbf{Fernando Uyaguari } received his MSc in Software Engineering, and PhD from Technical University of Madrid. His research interests include empirical software engineering and agile development methodologies, in particular test driven development. 
\newline

\clearpage

\hfill 
 \begin{floatingfigure}{1.2in}
 \includegraphics[width=1in,clip,keepaspectratio]{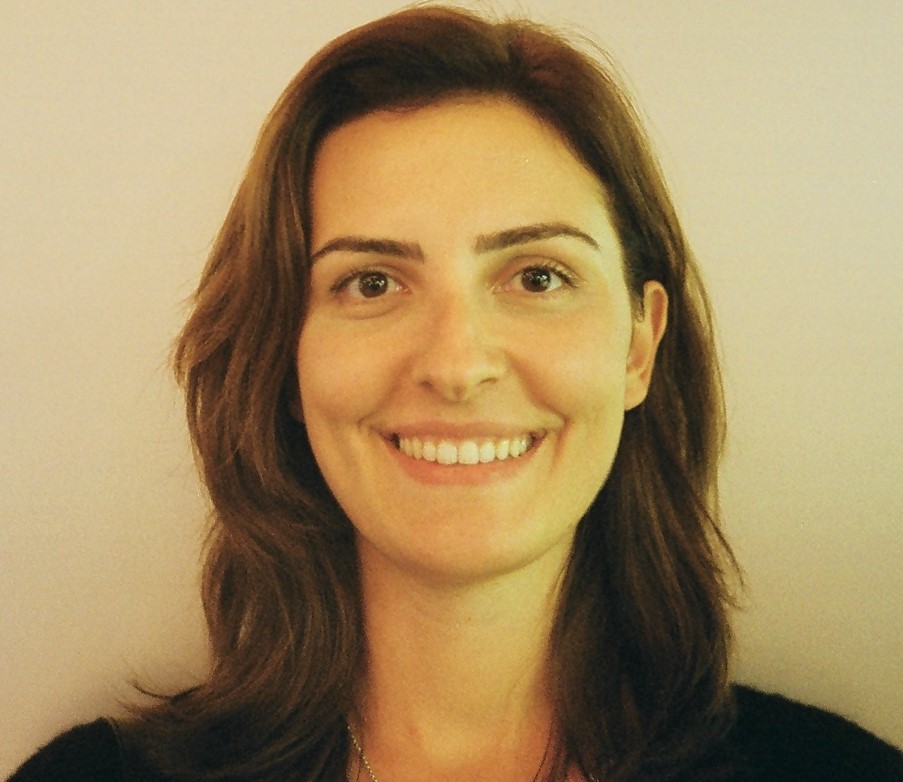}
 \end{floatingfigure}
 \noindent \textbf{Ayse Tosun} is an assistant professor at the Faculty of Computer and Informatics Engineering, Istanbul Technical University, Turkey. Prior to joining ITU, she worked as a post-doctoral research fellow at University of Oulu, Finland. She received her PhD in 2012, and MSc degree in 2008 from Department of Computer Engineering, Bogazici University, Turkey. Her research interests are empirical software engineering, more specifically mining software data repositories, software quality, measurement, process improvement, and applications of AI on building recommendation systems for software engineering. 
\newline


\hfill 
 \begin{floatingfigure}{1.2in}
 \includegraphics[width=1in,clip,keepaspectratio]{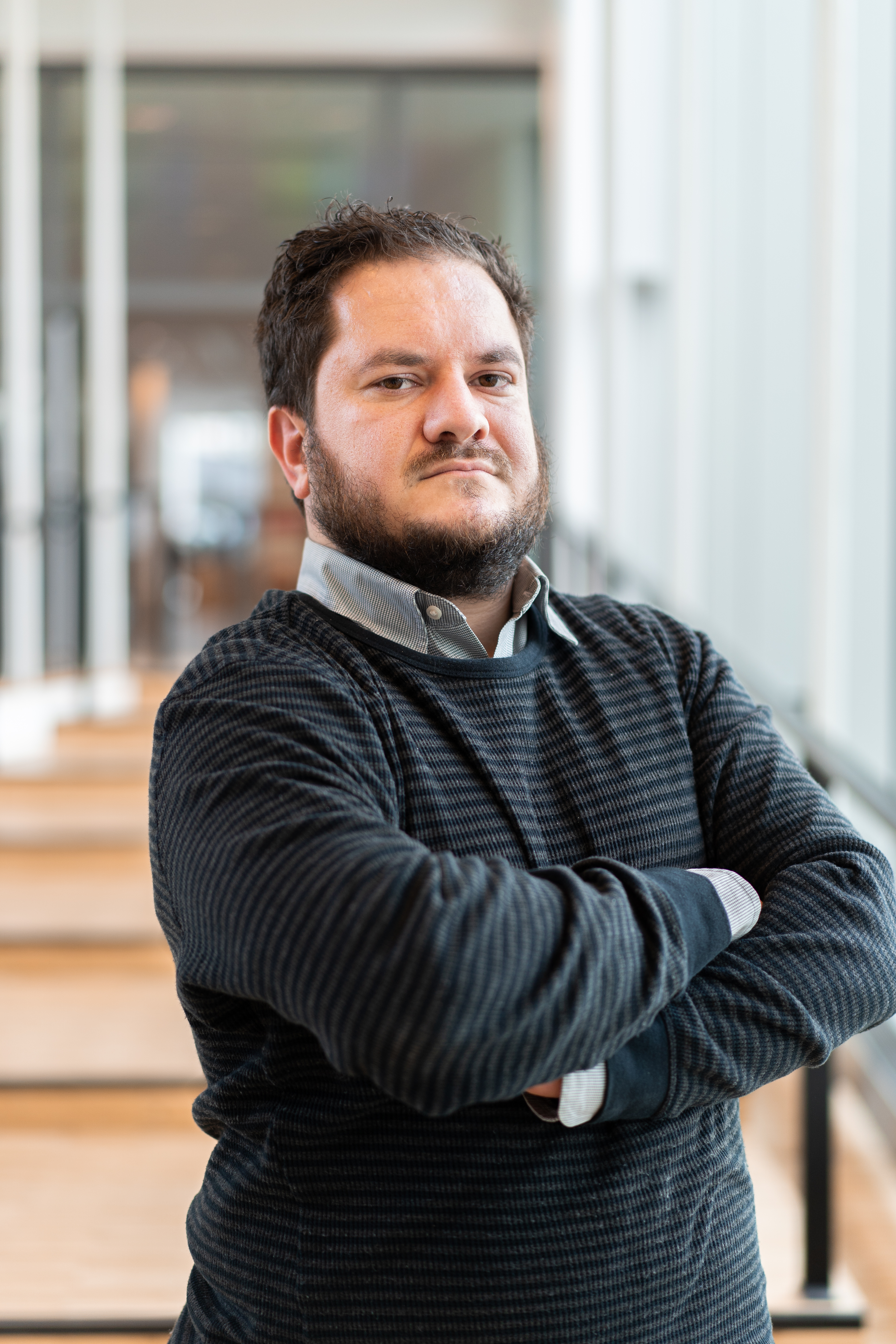}
 \end{floatingfigure}
 \noindent \textbf{Davide Fucci} is an Assistant Professor (tenure-track) in Software Engineering at the Blekinge Institute of Technology. His focus is on Empirical Software Engineering, Requirements Engineering, and Humas Aspects. During his Ph.D., Davide worked together with industry to evaluate and improve Agile software development methodologies, such as unit testing and test-driven development. Thereafter, he worked in the area of data-driven Requirements Engineering at the University of Hamburg. Davide regularly publishes in software engineering venues and occupies several positions in venues related to empirical software and requirements engineering communities. He is a member of the ACM and the IEEE Computer societies. 
\newline


\hfill 
 \begin{floatingfigure}{1.2in}
 \includegraphics[width=1in,clip,keepaspectratio]{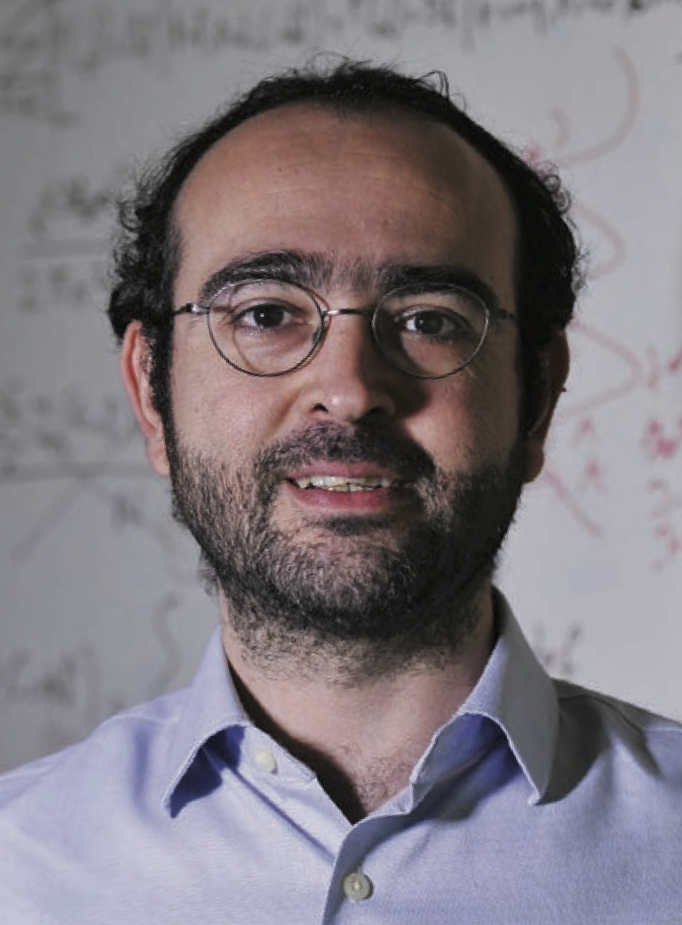}
 \end{floatingfigure}
 \noindent \textbf{Burak Turhan}, PhD (Boğaziçi University), is a Professor in the Faculty of Electrical Engineering and Information Technology at the University of Oulu and an Adjunct Professor in the Faculty of Information Technology at Monash University. His research focuses on empirical software engineering, software analytics, quality assurance and testing, human factors, and (agile) development processes. He is a Senior Associate Editor of Journal of Systems and Software, an Editorial Board Member of Empirical Software Engineering, Information and Software Technology, and Software Quality Journal, a Review Board member of IEEE Transactions on Software Engineering, a Senior Member of the ACM and IEEE, and a member of ACM SIGSOFT and IEEE Computer Society. For more information please visit: https://turhanb.net or contact him at turhanb@computer.org. 
\newline

\vspace{0.5cm}

\hfill 
 \begin{floatingfigure}{1.2in}
 \includegraphics[width=1in,clip,keepaspectratio]{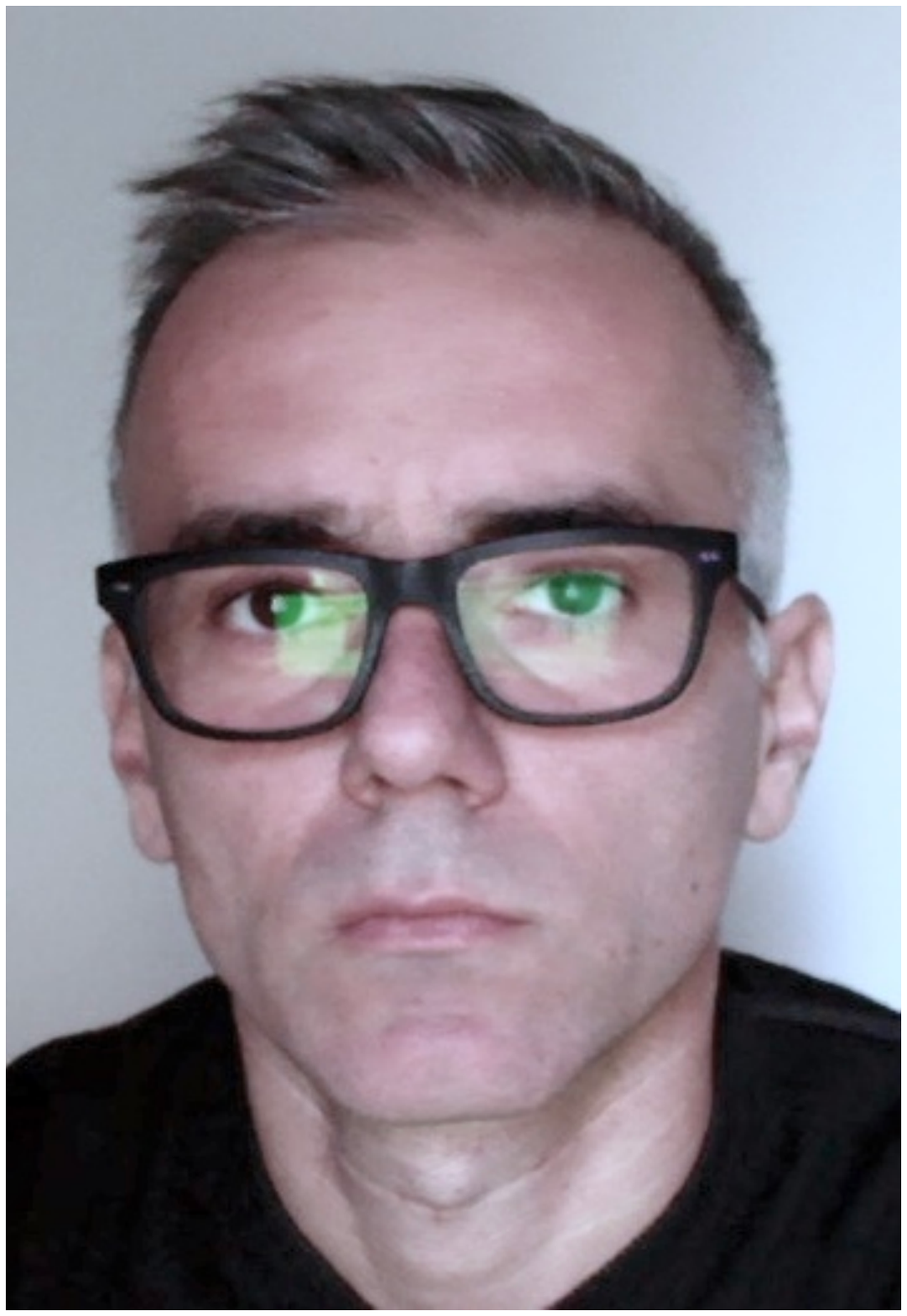}
 \end{floatingfigure}
 \noindent \textbf{Giuseppe Scanniello} received his Laurea andPh.D. degrees, both in Computer Science, from the University of Salerno, Italy, in 2001 and 2003, respectively. In 2006, he joined, as an AssistantProfessor, the Department of Mathematics andComputer Science at the University of Basilicata, Potenza, Italy. In 2015, he became an AssociateProfessor  at  the  same  university.  He  has  published more than 170 referred papers in journals, books, and conference proceedings. He serves on the organising of major international conferences  and  workshops  in  the  field  of  software  engineering.  GiuseppeScanniello  leads  both  the  group  and  the  laboratory  of  software  engineering at the University of Basilicata (BASELab). 
\newline


\hfill 
 \begin{floatingfigure}{1.2in}
 \includegraphics[width=1in,clip,keepaspectratio]{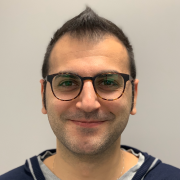}
 \end{floatingfigure}
 \noindent \textbf{Simone Romano} received the Ph.D. in Computer Science and Mathematics from the University of Salento (in collaboration with the University of Basilicata), Italy, in July 2018 under the supervision of prof. Giuseppe Scanniello. In September 2018, he joined the Department of Computer Science at the University of Bari as a postdoctoral research fellow. He has served in the organization and has been a program committee member of several international conferences/workshops. His research interests include: test-driven development, software refactoring, software testing, software maintenance, empirical software engineering, and human factors in software engineering 
\newline


\hfill 
 \begin{floatingfigure}{1.2in}
 \includegraphics[width=1in,clip,keepaspectratio]{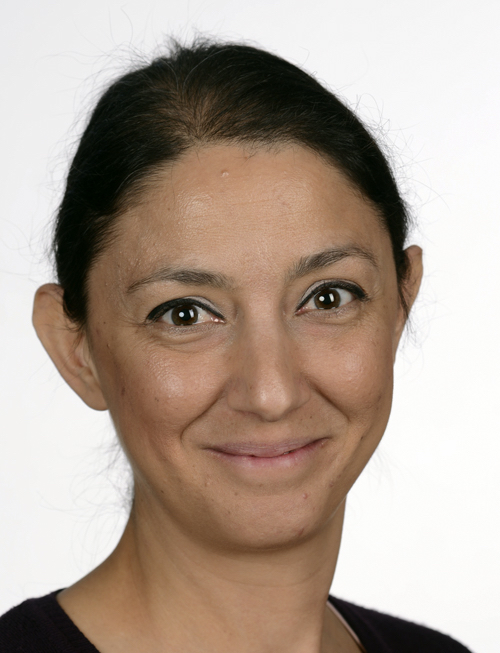}
 \end{floatingfigure}
 \noindent \textbf{Itir Karac} is a Ph.D. candidate and a researcher in the Empirical Software Engineering (M3S) research unit at the University of Oulu, Finland. She received her B.Sc. degree in Mathematics and her M.Sc. degree in Computer Engineering from Boğaziçi University, Turkey. Her research interests include empirical software engineering and software analytics. She is a member of IEEE. 
\newline

\clearpage

\hfill 
 \begin{floatingfigure}{1.2in}
 \includegraphics[width=1in,clip,keepaspectratio]{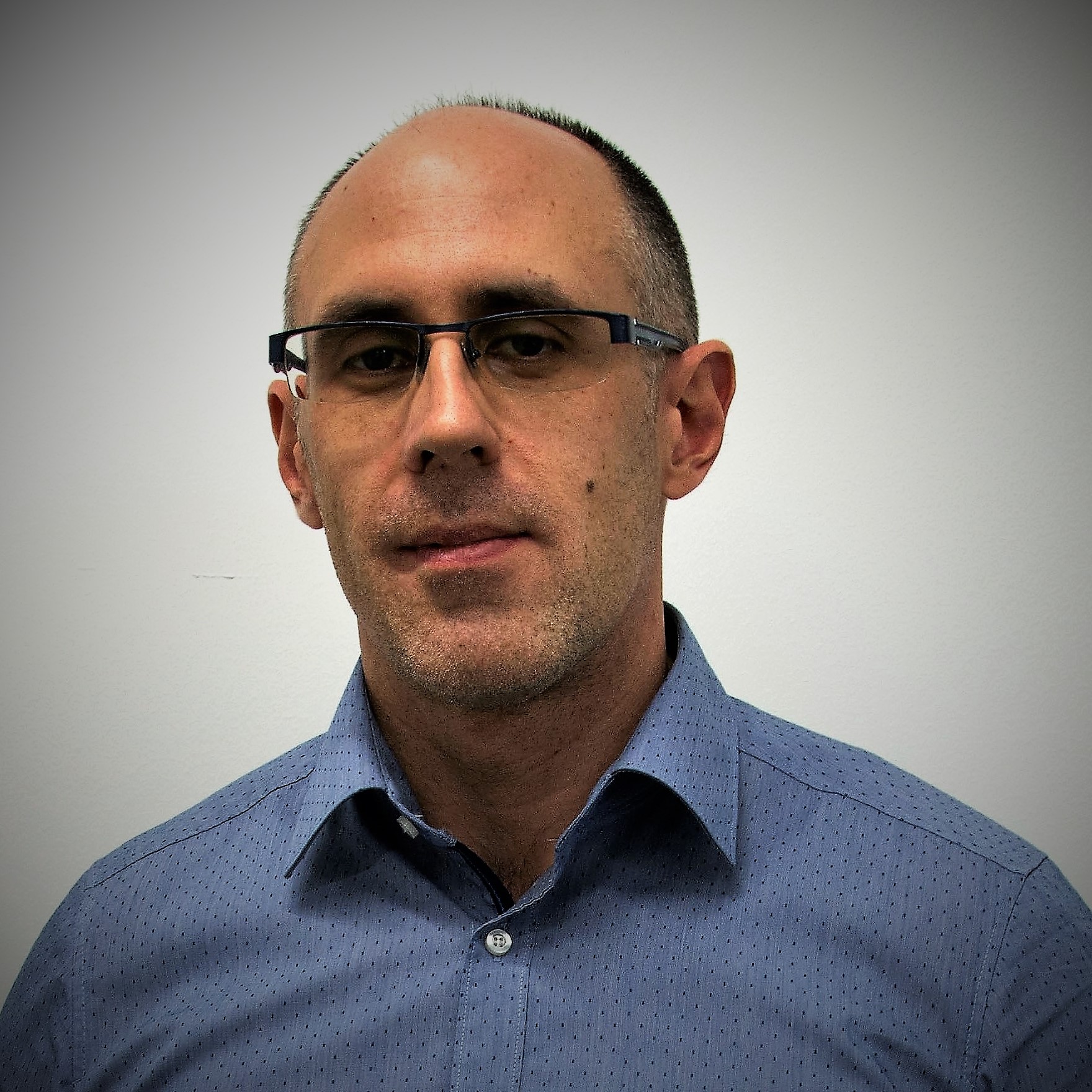}
 \end{floatingfigure}
 \noindent \textbf{Vladimir Mandić } is an assistant professor of SE at University of Novi Sad, Serbia. He received his PhD degree in Information Processing Science and SE from the University of Oulu, Finland, and M.Sc.E.E from the University of Novi Sad, Serbia. His areas of interest are software process improvement, empirical software engineering, goal-driven measurement approaches, technical debt and value-based software engineering. He is a member of the IEEE Computer Society. 
\newline


\hfill 
 \begin{floatingfigure}{1.2in}
 \includegraphics[width=1in,clip,keepaspectratio]{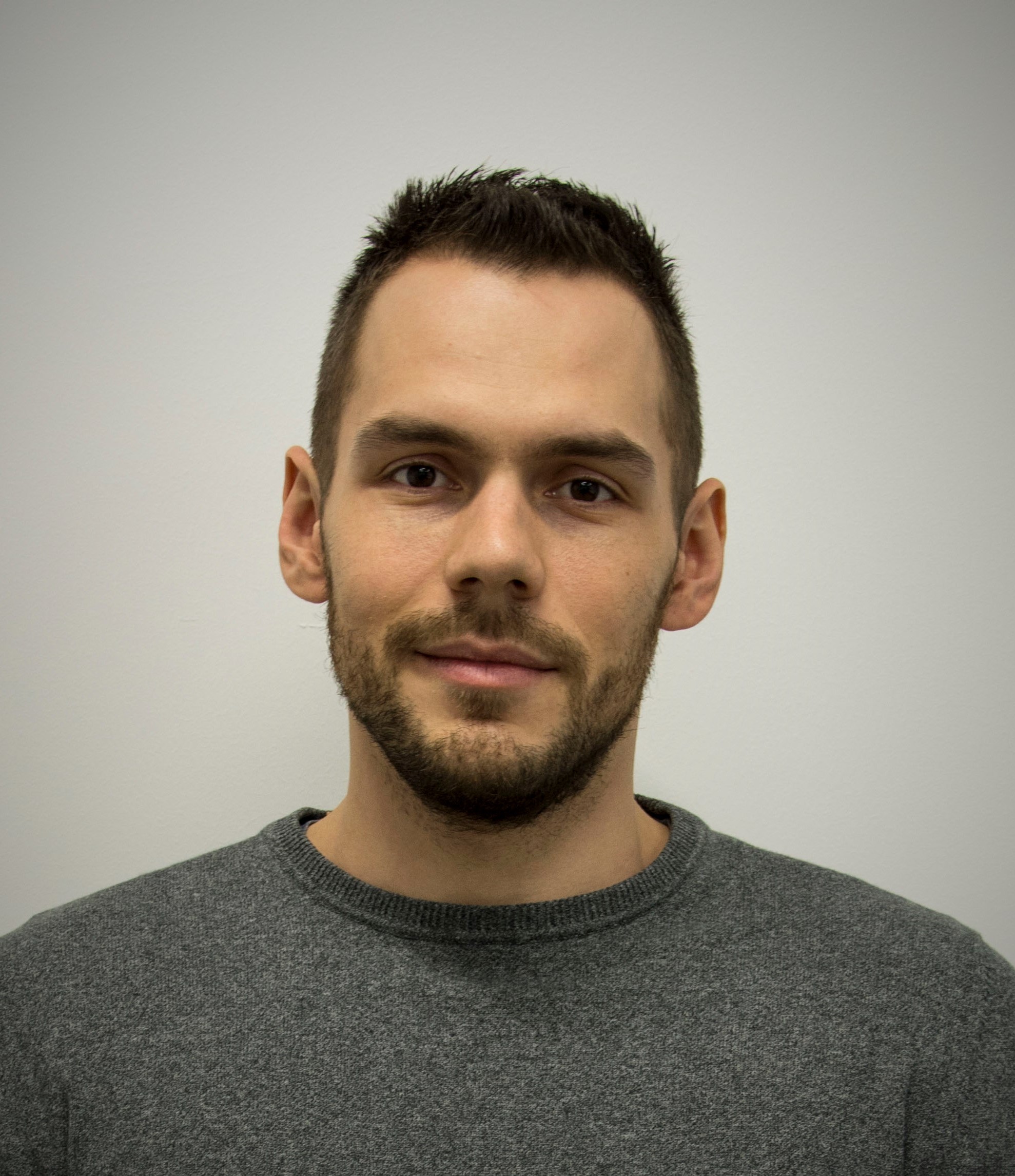}
 \end{floatingfigure}
 \noindent \textbf{Robert Ramač} is a teaching assistant of Software Engineering at the University of Novi Sad, Serbia, and a software developer at TIAC ltd. Currently he is enrolled as a PhD student at the University of Novi Sad, at the Faculty of Technical Sciences. He received two M.Sc. degrees at the University of Novi Sad, master of Information technologies and master of Engineering management. His areas of interest are: empirical software engineering, software development, technical debt, the improvement of the software development process.
\newline


\hfill 
 \begin{floatingfigure}{1.2in}
 \includegraphics[width=1in,clip,keepaspectratio]{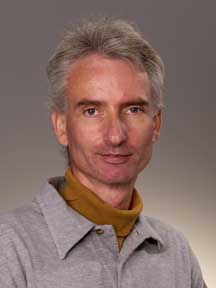}
 \end{floatingfigure}
 \noindent \textbf{Dietmar Pfahl} is Professor of Software Engineering at the University of Tartu, Estonia, and Adjunct Professor with the Schulich School of Engineering at the University of Calgary, Canada. He earned his PhD from TU Kaiserslautern, Germany in 2001. His research interests focus on data-driven decision support in software engineering and management. He has 140+ peer reviewed publications in top-ranked software engineering journals and conference proceedings. He is a senior member of both IEEE and ACM.  
\newline


\hfill 
 \begin{floatingfigure}{1.2in}
 \includegraphics[width=1in,clip,keepaspectratio]{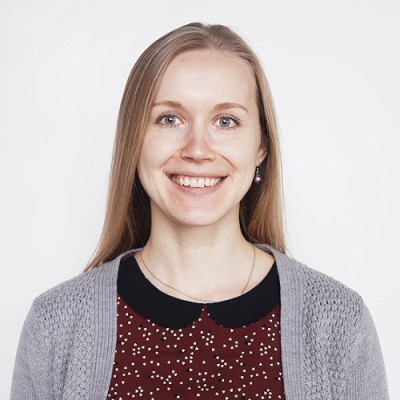}
 \end{floatingfigure}
 \noindent \textbf{Kerli Rungi} is Head of Learning and Growth at Testlio OY, Estonia. Prior to this, she worked at Playtech Estonia. Kerli has an MSc degree in Software Engineering from the University of Tartu, Estonia.
\newline

\clearpage

\hfill 
 \begin{floatingfigure}{1.2in}
 \includegraphics[width=1in,clip,keepaspectratio]{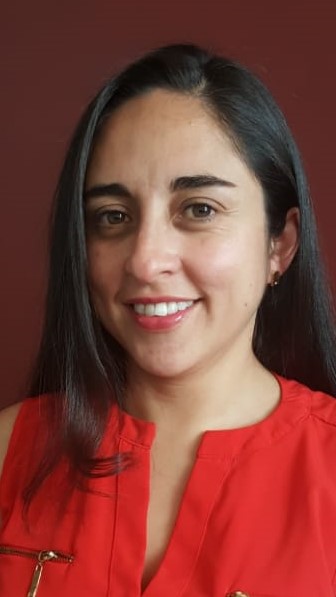}
 \end{floatingfigure}
 \noindent \textbf{Carolina Palomeque} is a systems engineer with a MS in IT Strategic Manager from Universidad de Cuenca, Ecuador. She has been working in the utilities industry for several years now in roles such as software engineer and project management. Her areas of interest include agile methodologies and empirical software engineering frameworks.
\newline

\vspace{1.8cm}

\hfill 
 \begin{floatingfigure}{1.2in}
 \includegraphics[width=1in,clip,keepaspectratio]{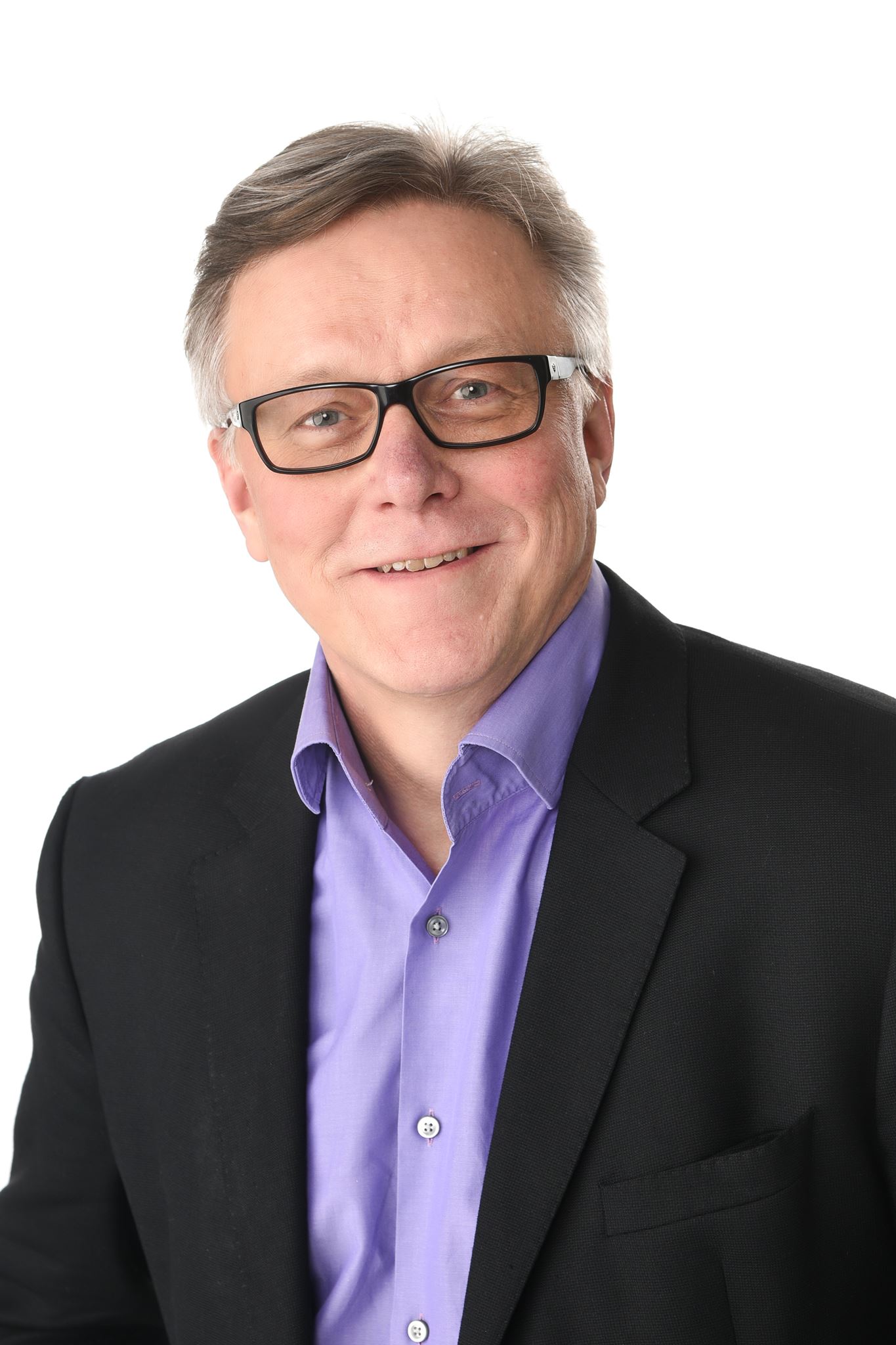}
 \end{floatingfigure}
 \noindent \textbf{Markku Oivo} is professor and head of the M3S research unit at the University of Oulu, Finland. He had visiting positions at the University of Maryland (1990-91), Schlumberger Ltd. (Paris 1994-95), Fraunhofer IESE (1999-2000), University of Bolzano (2014-15), and Universidad Polit\'ecnica de Madrid (2015).
\newline

\vspace{1.5cm}
    
\hfill     
 \begin{floatingfigure}{1.2in}
 \includegraphics[width=1in,clip,keepaspectratio]{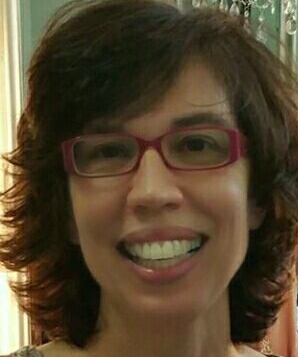}
 \end{floatingfigure}
 \noindent \textbf{Natalia Juristo} has been full professor of software engineering with the  School of Computer Engineering at the Technical University of Madrid, Spain, since 1997. She was awarded a FiDiPro (Finland Distinguished Professor Program) professorship at the University of Oulu, from January 2013 until June 2018. Natalia belongs to the editorial board of EMSE and STVR. In 2009, Natalia was awarded an honorary doctorate by Blekinge Institute of Technology in Sweden.

\end{document}